\documentclass{article}
\usepackage{graphicx} 

\usepackage{style}
\usepackage{authblk}

\title{Exact variance estimation for model-assisted survey estimators using U- and V-statistics}
\author[1]{Ameer Dharamshi}
\author[2]{Peter Gao}
\author[1,3]{Jon Wakefield}
\affil[1]{Department of Biostatistics, University of Washington}
\affil[2]{Department of Mathematics and Statistics, San José State University}
\affil[3]{Department of Statistics, University of Washington}

\newcommand{\blind}{0}

\begin{document}

\maketitle

\begin{abstract} 
Model-assisted estimation combines sample survey data with auxiliary information to increase precision when estimating finite population quantities. Accurately estimating the variance of model-assisted estimators is challenging:~the classical approach ignores uncertainty from estimating the working model for the functional relationship between survey and auxiliary variables. This approach may be asymptotically valid, but can underestimate variance in practical settings with limited sample sizes. In this work, we develop a connection between model-assisted estimation and the theory of U- and V-statistics. We demonstrate that when predictions from the working model for the variable of interest can be represented as a U- or V-statistic, the resulting model-assisted estimator also admits a U- or V-statistic representation. We exploit this connection to derive an improved estimator of the exact variance of such model-assisted estimators. The class of working models for which this strategy can be used is broad, ranging from linear models to modern ensemble methods. We apply our approach to the model-assisted estimator constructed with a linear regression working model, commonly referred to as the generalized regression estimator, show that it can be re-written as a U-statistic, and propose an estimator of its exact variance. We illustrate our proposal and compare it against the classical asymptotic variance estimator using household survey data from the American Community Survey.

\end{abstract}

\section{Introduction}

A common task in survey analysis is to estimate and perform inference on finite population quantities of interest using data collected with (possibly complex) surveys. For example, a national statistical office may be interested in estimating the average income in the previous year and over a set of groups (e.g., geographic areas) using data generated from a household survey. To answer such questions, a \emph{design-based} approach is frequently adopted. In the design-based framework, rather than modelling the outcome as a stochastic quantity, outcome variables are viewed as deterministic, and inference is carried out with respect to the randomization distribution induced by the sampling design, comparing the observed sample with all possible samples that could have been obtained from a fixed finite population of responses.

Formally, consider a population of interest $U=\{1,2,\dots,N\}$ where each unit $i\in U$ is associated with a fixed value of a target variable $y_i$. From this population, a subset of $n$ units, $S\subset U$, are sampled according to a pre-specified survey design, $p(S)$, which defines the probability of selecting every possible sample. Our objective is to use the observed sample $S$ to estimate various population summaries derived from $\{y_i\}_{i\in U}$. In this paper, we specifically focus on estimating the finite population average $a=\frac{1}{N}\sum_{i\in U} y_i$ though the ideas presented extend naturally to other estimands.

The foundational estimator in this context is the Horvitz-Thompson estimator \citep{horvitz1952generalization}. For every $i\in U$, let $I_i$ denote a sample inclusion indicator, that is, $I_i=1$ if and only if unit $i\in S$, and $I_i=0$ otherwise. Additionally, let $\pi_i=\E[I_i]$ and $\pi_{ij}=\E[I_iI_j]$ denote the first- and second-order inclusion probabilities, respectively. The Horvitz-Thompson estimator of $a$ is 
$$
\hat a_{HT}=\frac{1}{N}\sum_{i\in S}\frac{y_i}{\pi_i} = \frac{1}{N}\sum_{i\in U}\frac{y_iI_i}{\pi_i}.
$$ 
Assuming that $\pi_i>0$ for all $i\in U$, $\hat a_{HT}$ is design-unbiased in the sense that $E[\hat a_{HT}]=a$. Its variance is 
$$
\Var(\hat a_{HT})=\frac{1}{N^2}\sum_{i,j\in U}\Delta_{ij}\frac{y_i}{\pi_i}\frac{y_j}{\pi_j}
$$ 
where $\Delta_{ij}=\pi_{ij}-\pi_i\pi_j$, and if $\pi_{ij}>0$, it can be estimated with 
$$
\widehat\Var(\hat a_{HT})=\frac{1}{N^2}\sum_{i,j\in S}\Delta_{ij}\frac{y_i}{\pi_i}\frac{y_j}{\pi_j}\frac{1}{\pi_{ij}}=\frac{1}{N^2}\sum_{i,j\in U}\Delta_{ij}\frac{y_i}{\pi_i}\frac{y_j}{\pi_j}\frac{I_iI_j}{\pi_{ij}}.
$$

While elegant, the Horvitz-Thompson estimator can, unfortunately, be rather inefficient \citep{breidt2017review}. In modern practice, survey statisticians have access to auxiliary information collected on each unit in the population of interest and seek to exploit this additional information to improve efficiency. The \emph{model-assisted estimation} approach is one strategy that has attracted substantial interest in the literature. With model-assisted estimation, one begins by proposing a \emph{working model} for the functional relationship between the target variable $y_i$ and the auxiliary data vector $x_i$. The cornerstone of the model-assisted survey estimation framework is the \emph{difference estimator} \citep{cassel1976some, sarndal2003model}. The difference estimator predicts the outcome for all units in the population with some working model for $y_i$ given $x_i$ and corrects the average prediction with the weighted average of the observed residuals. Suppose we have access to some fixed (i.e., non-random) population-level working model, $m$. Then, the difference estimator is:
$$
\hat a_D = \frac{1}{N}\sum_{i\in U} m(x_i) + \frac{1}{N}\sum_{i\in S}\frac{y_i- m(x_i)}{\pi_i}.
$$

As $m$ is a \emph{population-level object}, the difference estimator inherits several attractive properties from the second Horvitz-Thompson term: 1. it is design-unbiased, 2. it is design-consistent, 3. it is asymptotically normal, and 4. under mild conditions, the variance of $\hat a_D$ can be estimated using the Horvitz-Thompson variance estimator on the model residuals \citep{sarndal2003model, breidt2017review}. Should the working model have predictive value, the variance of $\hat a_D$ will be lower than that of $\hat a_{HT}$ as the variance of the residuals will be lower than that of the outcome itself \citep{fuller2011sampling}.

Unfortunately, $m$ is rarely if ever available. Instead, it is typically replaced with a sample version, $\hat m$, fit to the observed pairs $\{(x_i,y_i)\}_{i\in S}$. This yields the \emph{model-assisted estimator}: 
\begin{equation}
\label{eq:mae}
\hat a = \frac{1}{N}\sum_{i\in U} \hat m(x_i) + \frac{1}{N}\sum_{i\in S}\frac{y_i- \hat m(x_i)}{\pi_i}.
\end{equation}

Since $\hat m$ is a \emph{sample-level object}, and thus is subject to sampling variability, analysing $\hat a$ is more challenging. Suppose the difference between $\hat m$ and $m$ is negligible (in the sense that the weighted predictions are asymptotically equivalent). In this case, $\hat a$ inherits asymptotic unbiasedness, asymptotic design-consistency, and asymptotic normality from $\hat a_D$ (see \cite{breidt2017review} for a comprehensive discussion of survey asymptotics and the asymptotic properties of $\hat a$). The following classical variance estimator can be used to estimate the asymptotic variance of $\hat a$, from consideration of the second term of \eqref{eq:mae} only: 
\begin{equation}
\label{eq:asymptotic}
\widehat\Var(\hat a)_{asy} = \sum_{i,j\in S}\frac{\Delta_{ij}}{\pi_{ij}}\frac{y_i-\hat m(x_i)}{\pi_i}\frac{y_j-\hat m(x_j)}{\pi_j}.
\end{equation}

Establishing this asymptotic equivalence for various candidate models has been a fruitful endeavour. The work of \cite{cassel1976some}, \cite{sarndal1980pi}, and \cite{sarndal2003model} explored the generalized regression (GREG) estimator in which $m$ is based on a linear regression model. More recent work has focused on machine learning models; see for example \cite{dagdoug2023model} for a discussion of using random forests as the working model.

In practical settings, however, it is often the case that $\hat m$ and $m$ are substantially different. This may be the result of a slow-to-converge model class, a complicated survey design, or a small-to-moderate sample size. When this happens, $\widehat\Var(\hat a)_{asy}$ will suffer from potentially severe downwards bias as it fails to account for the uncertainty stemming from $\hat m$. This issue has been discussed from both theoretical \citep{fuller2011sampling} and empirical \citep{mcconville2020tutorial} perspectives, though a solution remains elusive. 

In this paper, we propose a new approach to variance estimation for model-assisted estimators that is applicable when the working model can be expressed as either a U- or V-statistic. This is a broad class of working models, ranging from classical tools including linear models and B-splines to modern ensemble methods. Our approach is fundamentally different from the usual approach to variance estimation: rather than relying on asymptotic assumptions that ignore the uncertainty in the working model, we exploit the structure of the working model to target the \emph{exact variance}. Specifically, we show that the U- or V-statistic structure propagates to $\hat a$, thereby allowing one to use tools designed to estimate the exact variance of a U-statistic to estimate the exact variance of $\hat a$. By shifting the focus from the asymptotic variance to the exact variance, our proposal offers a pathway to more reliable finite sample inference.

The use of U-statistics in finite population survey sampling is not new. Building on the work of \cite{hoeffding}, \cite{Nandi1963OnTP} pioneered the extension of classical U-statistic results, including an exact variance expression, to study U-statistics where the data are drawn via simple random sampling without replacement from a finite population (i.e.,~they allow for a relaxation of independence). \cite{folsom} and \cite{williams} further extend and allow for more complicated survey designs. Chapter 2 of \cite{lee2019u} provides a compact summary of these ideas. 

Our work is inspired by, but different from, this line of inquiry. Rather than starting with a well-studied U-statistic and asking what happens if the data were generated by a survey, we start with a well-studied survey estimator and represent it as a U-statistic. This distinction is critical: it reduces the role of higher-order inclusion probabilities, paving the way for application beyond small toy examples. We demonstrate how to operationalise our theoretical developments using the generalized regression estimator as a motivating example.

The rest of this paper is organized as follows. In Section \ref{sec:uv}, we establish that for a subclass of working models, model-assisted estimators can be represented as U- and V-statistics, and use this connection to develop an expression for the exact variance of $\hat a$. In Section \ref{sec:greg}, we derive the U-statistic representation of the generalized regression estimator and subsequently, in Section \ref{sec:varest}, we outline a variance estimator based on this representation. In Section \ref{sec:sims}, we apply our approach to data from the American Community Survey. We conclude with a discussion in Section \ref{sec:discussion}. Proofs of technical results are deferred to the supplementary materials.

\section{Representing model-assisted estimators as U- and V-statistics}
\label{sec:uv}

\subsection{Connection to U- and V-statistics}

V-statistics are a class of estimators first studied by \cite{mises1947asymptotic}. Suppose we have data $z_j$ for $j=1,\dots n$, and a symmetric kernel function $h$ taking $k \le n$ arguments. A degree-$k$ V-statistic is 
$$
\frac{1}{n^k}\sum_{j_1=1}^n\dots\sum_{j_k=1}^nh(z_{j_1},\dots,z_{j_k}).
$$
V-statistics extend the notion of an average to higher-order functions: intuitively, they are the average of the kernel function $h$ evaluated at every possible subset of $k$ units drawn from the data $z_j$ (henceforth referred to as a $k$-subset), including subsets with repeated entries (i.e., they are drawn with replacement).

U-statistics are a closely related family of estimators first studied by \cite{hoeffding} in which instead of averaging over kernels evaluated on all possible $k$-subsets of $z_j$ drawn with replacement, we only average over all possible $k$-subsets of $z_j$ drawn without replacement. A degree-$k$ U-statistic is 
$$
\frac{1}{{n \choose k}}\sum_{j_1<j_2<\dots<j_k}h(z_{j_1},\dots,z_{j_k}).
$$

Returning to our setting of interest, let $z_j=(x_j,y_j,\pi_j,I_j)$. In the design-based framework, the only random object in $z_j$ is the inclusion indicator $I_j$. Suppose that we have a model $\hat m$ such that the prediction $\hat m(x_0)$ at a test point $x_0$ can be expressed as a degree-$k$ V-statistic with kernel $h(x_0|z_{j_1},\dots,z_{j_k})$ evaluated over the observed sample:
\begin{equation}
\label{eq:vstat}
\hat m(x_0)=\frac{1}{n^k}\sum_{j_1\in S}\dots\sum_{j_k\in S}h(x_0|z_{j_1},\dots,z_{j_k}).
\end{equation}

Many common working models can be expressed in this way. Anything that can be expressed as an average over the sample 
is a degree-$1$ V-statistic. Ensemble models, whether constructed from decision trees such as random forests, or from any other (weak) learner, can be expressed as higher degree V-statistics \citep{zhou2021v}. We will show that for these models, the V-statistic structure propagates to $\hat a$. 

Lemma \ref{lem:popvstat} states that $\hat m$ can alternatively be written as a V-statistic evaluated over the entire finite population, though with a slightly different kernel.

\begin{lemma}
\label{lem:popvstat}
Consider the working model $\hat m$ given in \eqref{eq:vstat}. It can be re-written as:
\begin{equation}
\label{eq:vstat2}
\hat m(x_0)=\frac{1}{N^k}\sum_{j_1\in U}\dots\sum_{j_k\in U}\frac{N^k}{n^k}h(x_0|z_{j_1},\dots,z_{j_k})\prod_{l=1}^k I_{j_l},
\end{equation}    
which is a degree-$k$ V-statistic with kernel $\frac{N^k}{n^k}h(x_0|z_{j_1},\dots,z_{j_k})\prod_{l=1}^k I_{j_l}$.
\end{lemma}

Now, returning to the model-assisted estimator, the next theorem proves that $\hat a$ can be written as a degree-$(k+1)$ V-statistic, though with a substantially more complicated kernel.

\begin{theorem}[Model-assisted estimators as V-statistics]
\label{thm:MAEvstat}
Consider the working model $\hat m$ given in \eqref{eq:vstat2} and define the model-assisted estimator of $a$ as $\hat a = \frac{1}{N}\sum_{i\in U} \hat m(x_i) + \frac{1}{N}\sum_{i\in S}\frac{y_i- \hat m(x_i)}{\pi_i}$. Then, $\hat a$ can be re-written as a degree-$(k+1)$ V-statistic:
\begin{equation}
\label{eq:vstat3}
\hat a = \frac{1}{N^{k+1}}\sum_{j_1\in U}\dots\sum_{j_{k+1}\in U}h^*(z_{j_1},\dots,z_{j_{k+1}})
\end{equation}
with kernel
$$
h^*(z_{j_1},\dots,z_{j_{k+1}})=\frac{1}{k+1}\sum_{p=1}^{k+1}\left[\frac{y_{j_p}I_{j_p}}{\pi_{j_p}} + \left(1-\frac{I_{j_p}}{\pi_{j_p}}\right)\frac{N^k}{n^k}h(x_{j_p}|\{z_{j_1},\dots,z_{j_{k+1}}\}\setminus \{z_{j_p}\})\prod_{l\in\{1,\dots,k+1\}\setminus \{p\}} I_{j_l}\right],
$$
where the notation $A\setminus B$ refers to the set of elements in $A$ that are not in $B$.
\end{theorem}

\begin{remark}
Notice that $h^*$ only depends on $y_{j_p}$ through the term $y_{j_p}I_{j_p}$. That is, $y_{j_p}$ only contributes to $h^*$ if it is observed. Consequently, $h^*$ can be evaluated \emph{for every} $(k+1)$-subset of the population, regardless of whether unobserved units are members of $\{z_{j_1},\dots,z_{j_{k+1}}\}$.
\end{remark}

\begin{remark}
Theorem \ref{thm:MAEvstat} can be easily adapted to other common targets of inference. For example, the standard model-assisted estimator of the population total can be expressed as a V-statistic with kernel $h^*_{\text{total}}(z_{j_1},\dots,z_{j_{k+1}})=Nh^*(z_{j_1},\dots,z_{j_{k+1}})$. While sightly more complicated, one can similarly express the estimator for the average within a domain (as is typically of interest in small-area estimation) as a V-statistic by appropriately inserting a domain inclusion indicator inside $h^*$. This yields 
\small
$$
h^*_{\text{SAE}}(z_{j_1},\dots,z_{j_{k+1}})=\frac{1}{k+1}\sum_{p=1}^{k+1}\left(\left[\frac{y_{j_p}I_{j_p}}{\pi_{j_p}} + \left(1-\frac{I_{j_p}}{\pi_{j_p}}\right)\frac{N^k}{n^k}h(x_{j_p}|\{z_{j_1},\dots,z_{j_{k+1}}\}\setminus \{z_{j_p}\})\prod_{l\in\{1,\dots,k+1\}\setminus \{p\}} I_{j_l}\right]I_{j_p}^{\text{A}}\right)
$$
\normalsize
where $I_{j_p}^{\text{A}}$ equals one if unit $j_p$ is a member of domain A and zero otherwise.
\end{remark}

Theorem \ref{thm:MAEvstat} is the link that will allow us to use classical and modern U- and V-statistics machinery to study $\hat a$. Variance estimation for V-statistics is, however, complicated due to the replication of data points in the kernel, and requires the asymptotic equivalence (as $N$ tends to infinity) of the target V-statistic and a U-statistic constructed with the same kernel \citep{lee2019u}. While Theorem \ref{thm:MAEvstat} may offer a new perspective on the asymptotic framework for model-assisted estimation, this implies that it cannot be used directly for our primary objective of estimating the exact variance of $\hat a$.

By contrast, variance estimation for U-statistics is fairly well-studied. \cite{hoeffding} derived an exact variance formula for U-statistics constructed with independent and identically distributed data and \cite{lee2019u} discusses several extensions that allow for either non-independent or non-identically distributed data. 

Recently, \cite{zhou2021v} demonstrated that any V-statistic can be recast as a U-statistic with a different kernel where the data replication is embedded \emph{inside} the kernel. Proposition \ref{prop:MAEustat} defines this representation for $\hat a$.

\begin{proposition}
\label{prop:MAEustat}
Consider the representation of $\hat a$ given in \eqref{eq:vstat3}. Then, $\hat a$ can be re-written as:
\begin{equation}
\label{eq:ustat}
\hat a = \frac{1}{{N\choose{k+1}}}\sum_{j_1 < \dots < j_{k+1}} \frac{{N\choose{k+1}}}{N^{k+1}} \sum_{s\in\mathcal{S}_{k+1}(\{j_1,\dots,j_{k+1}\})} \frac{1}{{{N-u(s)}\choose{k+1-u(s)}}}h^*(s)
\end{equation}
where $\mathcal{S}_{b}(B)$ refers to the set of all $b$-subsets drawn with replacement from the set $B$ and $u(s)$ counts the number of distinct elements in $s$. This is a degree-$(k+1)$ U-statistic with kernel
$$
h_U(z_{j_1},\dots,z_{j_{k+1}}) = \frac{{N\choose{k+1}}}{N^{k+1}} \sum_{s\in\mathcal{S}_{k+1}(\{j_1,\dots,j_{k+1}\})} \frac{1}{{{N-u(s)}\choose{k+1-u(s)}}}h^*(s).
$$
\end{proposition}

\begin{remark}
If the working model can be expressed as a U-statistic rather than a V-statistic, then one can define $\hat m(x_0)$ to be the out-of-bag prediction of $x_0$ and use similar steps as used in the proof of Theorem \ref{thm:MAEvstat} to directly write  $\hat a$ as a U-statistic, without resorting to an intermediate V-statistic step.
\end{remark}

\begin{remark}
Probability sample U-statistics have previously been considered in the literature; see for instance \cite{Nandi1963OnTP}, \cite{folsom}, \cite{williams}, and \cite{lee2019u}. These results generally operate by weighting individual kernel evaluations by the probability of observing the requisite units in the sample. As this quickly leads to intractable higher-order inclusion probabilities, these approaches have gained limited traction. Our construction is fundamentally different. Instead of placing the sampling indicators \emph{outside} the kernels and constructing weighted U-statistic over the sample, we embed the sampling indicators \emph{inside} the kernels, enabling an unweighted U-statistic over the population, which reduces the role of higher-order inclusion probabilities. 
\end{remark}

\subsection{Variance estimation}
\label{subsec:hdecomp}

The representation of $\hat a$ in Proposition \ref{prop:MAEustat} is useful as it unlocks variance estimation tools specifically designed to exploit the symmetric structure of U-statistics. 
Our approach centres around the H-decomposition, a tool frequently used in the study of U-statistics, that re-writes a degree-$k$ U-statistic as the sum of $k$ lower degree U-statistics \citep{hoeffding, lee2019u, dehling2006limit, han2018inference}. We construct the following H-decomposition, which is related to the constructions in \cite{lee2019u} and \cite{han2018inference}, but has convenient properties even when the data are neither independent nor identically distributed (i.e.,~under any sampling design).

Following the notation in Proposition \ref{prop:MAEustat}, first define $\theta_{j_1,\dots,j_{k+1}}=\E[h_U(z_{j_1},\dots,z_{j_{k+1}})]$ where the expectation is taken with respect to $I_{j_1},\dots,I_{j_{k+1}}$. Then, for $1\le c \le k$, let 
$$
\theta'_{c;j_1,\dots,j_c}=\frac{1}{{{N-c}\choose{k+1-c}}}\sum_{t\in\mathcal{T}_{k+1-c}(U\setminus\{j_1,\dots,j_c\})}\theta_{j_1,\dots,j_c,t},
$$
where $\mathcal{T}_{b}(B)$ refers to the set of all $b$-subsets drawn without replacement from the set $B$. 

Similarly, define the conditional expectation 
$$
\phi_{c;j_1,\dots,j_{k+1}}(z_{j_1},\dots,z_{j_c})=\E[h_U(z_{j_1},\dots,z_{j_{k+1}})|z_{j_1},\dots,z_{j_c}]
$$ 
where the expectation is taken with respect to $I_{j_{c+1}},\dots,I_{j_{k+1}}$ and $1\le c \le k$. Then, let 
$$
\phi'_{c;j_1,\dots,j_c}(z_{j_1},\dots,z_{j_c})=\frac{1}{{{N-c}\choose{k+1-c}}}\sum_{t\in\mathcal{T}_{k+1-c}(U\setminus\{j_1,\dots,j_{k+1}\})}\phi_{c;j_1,\dots,j_c,t}(z_{j_1},\dots,z_{j_c}).
$$
The quantities $\theta'_{c;j_1,\dots,j_c}$ and $\phi'_{c;j_1,\dots,j_c}(z_{j_1},\dots,z_{j_c})$ are understood as the average of all kernel means that include the subset $\{j_1,\dots,j_c\}$ and the average of the same kernels conditioned on the event that the subset $\{j_1,\dots,j_c\}$ is fixed, respectively.

Next, define the functions 
\begin{align*}
h_{1;j_1}(z_{j_1}) &= \phi'_{1;j_1}(z_{j_1}) - \theta'_{1;j_1}, \\
h_{c;j_1,\dots,j_c}(z_{j_1},\dots,z_{j_c}) &= \phi'_{c;j_1,\dots,j_c}(z_{j_1},\dots,z_{j_c}) - \sum_{d=1}^{c-1}\left[\sum_{\substack{\{t_1,\dots,t_d\}\in \\ \mathcal{T}_d(\{j_1,\dots,j_c\})}}h_{d;t_1,\dots,t_d}(z_{t_1},\dots,z_{t_d})\right] - \theta'_{c;j_1,\dots,j_c}, \\
h_{k+1;j_1,\dots,j_{k+1}}(z_{j_1},\dots,z_{j_{k+1}}) &= h_U(z_{j_1},\dots,z_{j_{k+1}}) - \sum_{d=1}^{k}\left[\sum_{\substack{\{t_1,\dots,t_d\}\in \\ \mathcal{T}_d(\{j_1,\dots,j_{k+1}\})}}h_{d;t_1,\dots,t_d}(z_{t_1},\dots,z_{t_d})\right] - \theta_{j_1,\dots,j_{k+1}}.
\end{align*}

\begin{remark}
\label{rem:hfunctions}
The above are series of nested functions built from differences between averages of conditional expectations and their corresponding expectations. It follows from this structure that
$$
\E[h_{1;j_1}(z_{j_1})] = 0; \quad \E[h_{c;j_1,\dots,j_c}(z_{j_1},\dots,z_{j_c})] = 0\; \textrm{ for }\; 2\le c \le k; \; \textrm{and}\quad \E[h_{k+1;j_1,\dots,j_{k+1}}(z_{j_1},\dots,z_{j_{k+1}})] = 0.
$$
\end{remark}

Finally, the H-decomposition of $\hat a$ is given by
\begin{equation}
\label{eq:hdecomp}
\hat a = \theta + \sum_{c=1}^{k+1}{{k+1}\choose c}H_c
\end{equation}
where $\theta=\frac{1}{{N\choose{k+1}}}\sum_{j_1<j_2<\dots<j_{k+1}}\theta_{j_1,\dots,j_{k+1}}$ and $H_c$ is the degree-$c$ U-statistic constructed with the kernel $h_{c;j_1,\dots,j_c}(z_{j_1},\dots,z_{j_c})$. It then follows that the variance of $\hat a$ can be written as 
\begin{equation}
\label{eq:Hvar}
\Var(\hat a) = \sum_{c=1}^{k+1}{{k+1}\choose c}^2\tau_c + 2\sum_{1\le c < d \le k+1}{{k+1}\choose c}{{k+1}\choose d}\omega_{cd}
\end{equation}
where $\tau_c=\Var(H_c)$ and $\omega_{cd}=\Cov(H_c,H_d)$.

The utility of the variance expression in \eqref{eq:Hvar} is twofold. First, when the units are independent (i.e.,~under Bernoulli or Poisson sampling), the $H_c$ terms in \eqref{eq:hdecomp} are orthogonal and thus the $\omega$ terms in \eqref{eq:Hvar} are all zero \citep{lee2019u, van2000asymptotic}. In practice, it will often be reasonable to ignore the $\omega$ terms, even for non-independent sampling designs, as they are dominated by the $\tau$ terms. A second benefit is that it is much easier to estimate each $\tau_c$ term individually as opposed to attempting to estimate $\Var(\hat a)$ directly. 

The details of how to estimate each $\tau_c$ depends on the exact form of $h_U$, which in turn depends on the working model of choice. For the remainder of this paper, we focus on the commonly used generalized regression (GREG) estimator for which $m$ is a linear model (see \cite{breidt2017review} for a recent review). We first derive the form of $h_U$ for this case and then propose an estimator for each $\tau_c$ term, and therefore $\Var(\hat a)$.

\section{Revisiting the generalized regression estimator}
\label{sec:greg}

The GREG estimator begins with the following working model for the outcome:
$$
y_i = x_i^\top\beta + \epsilon_i,
$$
where $\epsilon_i$ are independent noise terms with $\Var(\epsilon_i)=\sigma^2<\infty$. Following the usual workflow for model-assisted estimators, we then estimate $\beta$ from the observed sample with weighted least squares using the survey weights, yielding the following working model:
$$
\hat m(x_i) = x_i^\top Q\sum_{j\in S} \frac{x_j y_j}{\pi_j}.
$$
where $Q^{-1}=E\left[x x^\top\right]=\sum_{i\in U}x_i x_i^\top$. For the purposes of this paper, since we assume that we observe $\{x_i\}_{i\in U}$, $Q$ is computed exactly, rather than estimated from the sample. (Technically, we only require knowledge of the first two moments of $x_i$, however, for other working models, all of $\{x_i\}_{i\in U}$ may be required.)

Using $\hat m(x_i)$, we estimate $\hat a$ with \eqref{eq:mae}. Since $\hat m(x_i)$ is a sum over the sample, it can be written as a degree-$1$ V-statistic and thus $\hat a$ can be expressed as a degree-$2$ U- or V-statistic.

The next result derives this new representation of $\hat a$ as a U-statistic. Following the process outlined in Section \ref{sec:uv}, we first re-write $\hat a$ as a double summation over the population, and then symmetrize to recover $h^*$. This yields the V-statistic representation given in Proposition \ref{prop:Vgreg}. 

\begin{proposition}
\label{prop:Vgreg}
The GREG estimator of $\hat a$ admits a representation as a degree-2 V-statistic with kernel
$$
h^*(z_i,z_j)=\frac{1}{2}\left[\left(1+N\left(\frac{1}{N}t_x^\top-\frac{I_j}{\pi_j}x_j^\top\right)Qx_i\right)\frac{y_iI_i}{\pi_i}+\left(1+N\left(\frac{1}{N}t_x^\top-\frac{I_i}{\pi_i}x_i^\top\right)Qx_j\right)\frac{y_jI_j}{\pi_j}\right]
$$
where $t_x=\sum_{i\in U}x_i$.
\end{proposition}

Next, we aggregate kernels with common terms; when $k+1$ is small (here $2$), this is a tractable endeavour. Indeed, the form of $h_U$ (defined in Proposition \ref{prop:MAEustat}) is
\begin{align*}
h_U(z_i,z_j)&=\frac{{N\choose2}}{N^2}\left[h^*(z_i,z_j) + h^*(z_j,z_i) + \frac{1}{N-1}h^*(z_i,z_i) + \frac{1}{N-1}h^*(z_j,z_j)\right] \\
&=\frac{N-1}{2N}\left[2h^*(z_i,z_j) + \frac{1}{N-1}h^*(z_i,z_i) + \frac{1}{N-1}h^*(z_j,z_j)\right].
\end{align*}

Under the design-based framework the randomness in $h_U(z_i,z_j)$ is derived exclusively from the two inclusion indicators, $I_i$ and $I_j$, so we can further simplify into the four cases corresponding to whether the units $i$ and $j$ are observed:
\begin{equation}
\label{eq:hUcases}
h_U(z_i,z_j) = \begin{cases}
\frac{N-1}{2N}\left[2h^*(z_i,z_j) + \frac{1}{N-1}h^*(z_i,z_i) + \frac{1}{N-1}h^*(z_j,z_j)\right] & I_i=I_j=1 \\
\frac{N-1}{2N}\left[\left(1+t_x^\top Qx_i\right)\frac{y_i}{\pi_i} + \frac{1}{N-1}h^*(z_i,z_i)\right] & I_i=1; I_j=0 \\
\frac{N-1}{2N}\left[\left(1+t_x^\top Qx_j\right)\frac{y_j}{\pi_j} + \frac{1}{N-1}h^*(z_j,z_j)\right] & I_i=0; I_j=1 \\
0 & I_i=I_j=0
\end{cases}.
\end{equation}
In further recognition of the role of the inclusion indicators, we will refer to the first three cases as $h_U(I_i=1,I_j=1)$, $h_U(I_i=1,I_j=0)$, and $h_U(I_i=0,I_j=1)$, respectively (the fourth case is omitted as it is constant at zero).
This expression will prove useful in the next section where we develop an estimator of $\Var(\hat a)$. 
\section{Exact variance estimation for generalized regression estimators}
\label{sec:varest}

In this section, we develop a procedure for estimating $\Var(\hat a)$ where $\hat a$ is the GREG estimator of the population average of $y$ using the U-statistic representation developed in \eqref{eq:hUcases}. Simplifying \eqref{eq:Hvar} to the setting where $k+1=2$, we have 
$$
\Var(\hat a) = 4\tau_1 + \tau_2 + 4\omega_{12}.
$$

We develop an estimator for $\Var(\hat a)$ under Poisson sampling designs since, as discussed in Section \ref{subsec:hdecomp}, the independence of the sample inclusion indicators $I_i$ implies that $\omega_{12}=0$. We can therefore focus on developing an approach to estimate the two $\tau$ terms.
Deriving variance estimators for Poisson sampling and then exploring their reliability in more complicated designs is a frequently used tactic; see for example \cite{yung1996jackknife} and \cite{valliant2002variance}.  As we will see in Section \ref{sec:sims}, assuming that $\omega_{12}$ is negligible may be reasonable even for non-Poisson sampling schemes.

We begin with $\tau_1$. From Section \ref{sec:uv}, $\tau_1$ is defined as
\begin{equation}
\label{eq:varH1}
\tau_1=\Var(H_1)=\frac{1}{N^2}\sum_{i=1}^N\Var(h_{1;i}(z_i))=\frac{1}{N^2}\sum_{i=1}^N\Var(\phi'_{1;i}(z_i)-\theta'_{1;i})=\frac{1}{N^2}\sum_{i=1}^N\E\left[(\phi'_{1;i}(z_i)-\theta'_{1;i})^2\right],
\end{equation}
where the last equality follows from the fact that $E[\phi'_{1;i}(z_i)-\theta'_{1;i}] = 0$.

Recall that $z_i$ takes on one of two values: $z_i=\{x_i,y_i,\pi_i,I_i=0\}$ or $z_i=\{x_i,y_i,\pi_i,I_i=1\}$. It follows that $\phi'_{1;i}(z_i)$, a function of $z_i$, also takes on one of two values. We will refer to these cases with $\phi'_{1;i}(I_i=0)$ and $\phi'_{1;i}(I_i=1)$, respectively. Next, using the fact that $\phi'_{1;i}(z_i)$ equals $\phi'_{1;i}(I_i=1)$ with probability $\pi_i$ (and $\phi'_{1;i}(I_i=0)$ otherwise) and that $E[\phi'_{1;i}(z_i)-\theta'_{1;i}] = 0 \implies  [\phi'_{1;i}(I_i=0)-\theta'_{1;i}](1-\pi_i)=-[\phi'_{1;i}(I_i=1)-\theta'_{1;i}]\pi_i$, observe that \eqref{eq:varH1} further simplifies as
\begin{align}
\tau_1&=\frac{1}{N^2}\sum_{i=1}^N\E\left[(\phi'_{1;i}(z_i)-\theta'_{1;i})^2\right] \nonumber \\ 
&= \frac{1}{N^2}\sum_{i=1}^N\left[(\phi'_{1;i}(I_i=0)-\theta'_{1;i})^2(1-\pi_i)+(\phi'_{1;i}(I_i=1)-\theta'_{1;i})^2\pi_i\right] \nonumber  \\
&= \frac{1}{N^2}\sum_{i=1}^N\left[(\phi'_{1;i}(I_i=1)-\theta'_{1;i})^2\frac{\pi_i^2}{1-\pi_i}+(\phi'_{1;i}(I_i=1)-\theta'_{1;i})^2\pi_i\right] \nonumber  \\
&=\frac{1}{N^2}\sum_{i=1}^N\left(\phi'_{1;i}(I_i=1)-\theta'_{1;i}\right)^2\frac{\pi_i}{1-\pi_i} \label{eq:tau1}
\end{align}

To estimate $\tau_1$ using \eqref{eq:tau1}, we require estimates of $\phi'_{1;i}(I_i=1)$ and $\theta'_{1;i}$. Using the notation defined in $\eqref{eq:hUcases}$, it follows from the law of total expectation that $\theta_{i,j}=h_U(I_i=1,I_j=1)\pi_{i}\pi_{j}+h_U(I_i=1,I_j=0)\pi_i(1-\pi_{j})+h_U(I_i=0,I_j=1)(1-\pi_{i})\pi_j$. Similarly, $\phi_{1;i,j}(I_i=1)=h_U(I_i=1,I_j=1)\pi_j + h_U(I_i=1,I_j=0)(1-\pi_j)$. Notice that $\theta_{i,j}$ and $\phi_{1;i,j}(I_i=1)$ are functions of $y_i$ through the $h_U$ terms. This will preclude estimation of $\phi'_{1;i}(I_i=1)$ and $\theta'_{1;i}$ for unobserved units. Instead of estimating all $N$ terms of \eqref{eq:tau1}, we focus on units $i\in S$ and construct the estimators $\hat\phi'_{1;i}(I_i=1)=\frac{1}{N-\frac{1}{\pi_i}}\sum_{j\ne i}\phi_{1;i,j}(I_i=1)\frac{I_j}{\pi_j}$ and $\hat\theta'_{1;i}=\frac{1}{N-\frac{1}{\pi_i}}\sum_{j\ne i}\theta_{i,j}\frac{I_j}{\pi_j}$. We then estimate $\tau_1$ with the following Horvitz-Thompson estimator:
\begin{equation}
\label{eq:tau1hat}
\hat\tau_1 = \frac{1}{N^2}\sum_{i\in S}\left(\hat\phi'_{1;i}(I_i=1)-\hat\theta'_{1;i}\right)^2\frac{1}{1-\pi_i}.
\end{equation}

\begin{remark}
\label{rem:ij}
Equations \eqref{eq:varH1} and \eqref{eq:tau1} are related to the infinitesimal jackknife variance estimator for U- and V-statistics \citep{efron2014estimation}, specifically, the balanced variance estimation method (BM) representation of the infinitesimal jackknife proposed by \cite{zhou2021v}. We expand on this connection in Supplement \ref{app:IJ}.
\end{remark}

The problem of estimating the variance of conditional means such as in $\tau_1$ is well-studied, and common estimation algorithms routinely suffer from upwards bias in finite samples \citep{sun2011efficient, wager2014confidence, zhou2021v}. In our case, the process of estimating $\hat\phi'_{1;i}(I_i=1)$ and $\hat\theta'_{1;i}$ induces unwanted excess variance. 
Fortunately, this excess variance can be analysed and excised. The following proposition states the form of the bias, which we subsequently propose a correction for.

\begin{proposition}
\label{prop:tau1bias}
The estimate of $\tau_1$ given in \eqref{eq:tau1hat} is biased upwards with magnitude
\begin{align*}
b_{\hat\tau_1}&=E[\hat\tau_1]-\tau_1=\frac{1}{N^2}\sum_{i=1}^N\Var\left(\hat\phi_{1;i}'(I_i=1)-\hat\theta'_{1;i}\right)\frac{\pi_i}{1-\pi_i} \\
&=\frac{1}{N^2}\sum_{i=1}^N\left(\frac{1}{\left(N-\frac{1}{\pi_i}\right)^2}\sum_{j\ne i}\frac{1-\pi_j}{\pi_j}\cdot\left(\phi_{1;i,j}(I_i=1)-\theta_{i,j}\right)^2\right)\frac{\pi_i}{1-\pi_i}.
\end{align*}
\end{proposition}

Proposition \ref{prop:tau1bias} says that the bias in $\hat\tau_1$ is simply the sum of the (scaled) variances of the estimates $\hat\phi'_{1;i}(I_i=1)-\hat\theta'_{1;i}$. Moreover, the second expression for $b_{\hat\tau_1}$ exploits the fact that each $\hat\phi'_{1;i}(I_i=1)-\hat\theta'_{1;i}$ is a Horvitz-Thompson estimator to derive a formula for the bias correction. 

To estimate $b_{\hat\tau_1}$, we consider the following weighted estimator
$$
\hat b_{\hat\tau_1}=\frac{1}{N^2}\sum_{i\in S}\left(\frac{1}{\left(N-\frac{1}{\pi_i}\right)^2}\sum_{j\in S\setminus\{i\}}\frac{1-\pi_j}{\pi_j^2}\cdot\left(\phi_{1;i,j}(I_i=1)-\theta_{i,j}\right)^2\right)\frac{1}{1-\pi_i}
$$
which then leads to the final estimator for $\tau_1$: $\hat\tau_{1,BC}=\hat\tau_1 - \hat b_{\hat\tau_1}$.

Turning to $\tau_2$, the following proposition describes an approximation that we will rely on for estimation. 

\begin{proposition}
\label{prop:tau2}
$\tau_2=\Var(H_2)$ is well approximated by
$$
\tau_2\approx\frac{4}{N^2(N-1)^2}\left(\underbrace{\sum_{1\le i<j\le N}\E\left[\left(h_U(z_i,z_j)-\phi'_{1;i}(z_i)-\phi'_{1;j}(z_j)+a\right)^2\right]}_{\tau_{2a}}-\underbrace{\sum_{1\le i<j\le N}\left(\theta_{i,j}-\theta'_{1;i}-\theta'_{1;j}+a\right)^2}_{\tau_{2b}}\right)
$$
under the assumption that
$$
\frac{4}{N^2(N-1)^2}\sum_{\substack{1\le i_1<j_2\le N, 1\le i_2<j_2\le N \\ \{i_1,j_1\}\ne\{i_2,j_2\}}}\E\left[h_{2;i_1,j_1}(z_{i_1},z_{j_1})h_{2;i_2,j_2}(z_{i_2},z_{j_2})\right]
$$
is small.
\end{proposition}

The approximation in Proposition \ref{prop:tau2} is obtained by expanding $\Var(H_2)$ and discarding the cross terms as negligible. This is a simplifying assumption grounded in the fact that they are a second order contribution to a second order term. To estimate $\tau_2$ using this approximation, we make a further assumption that $\tau_{2b}$ is negligible and focus on $\tau_{2a}$. In Supplement \ref{app:tau2b}, we provide a strategy for estimating $\tau_{2b}$ and empirically verify that indeed it represents a negligible contribution to $\Var(\hat a)$ as expected. 

Returning to $\tau_{2a}$, consider the following plug-in estimator where we plug-in $\hat\phi'_{1;i}(I_i=1)$ for $\phi'_{1;i}(z_i)$ when $i\in S$ and $\hat\phi'_{1;i}(I_i=0)$ otherwise:
\begin{equation}
\label{eq:tau2ahat}
\hat\tau_{2a}=\sum_{1\le i<j\le N}\left(h_U(z_i,z_j)-\hat\phi'_{1;i}(z_i)-\hat\phi'_{1;j}(z_j)+\hat a\right)^2.
\end{equation}
This estimator is convenient for two reasons. First, it is tractable in the sense that the expression $h_U(z_i,z_j)-\hat\phi'_{1;i}(z_i)-\hat\phi'_{1;j}(z_j)+\hat a$ can be computed for all pairs of units in the population. Second, it recycles terms computed while estimating $\hat\tau_1$ which is useful computationally. 

The usage of $\hat\phi'_{1;i}(z_i)$ here does, however, imply that $\hat\tau_{2a}$ is subject to an upwards bias as well. Following the same logic as used in Proposition \ref{prop:tau1bias}, the magnitude of the bias can be well approximated by $b_{\hat\tau_{2a}}=\sum_{1\le i < j \le N}\Var\left(\hat\phi_{1;i}'(z_i)+\hat\phi_{1;j}'(z_j)\right)$ which is again the sum of variances of Horvitz-Thompson estimators. We therefore estimate $b_{\hat\tau_{2a}}$ with 
$$
\hat b_{\hat\tau_{2a}} =  \sum_{i,j\in S}\left(\left(\frac{2N-\frac{1}{\pi_i}-\frac{1}{\pi_j}}{\left(N-\frac{1}{\pi_i}\right)\left(N-\frac{1}{\pi_i}\right)}\right)^2\sum_{k\in S}\frac{1-\pi_k}{\pi_k^2}\cdot\left(\phi_{1;i,k}(z_i)I_{k\ne i}+\phi_{1;j,k}(z_j)I_{k\ne j}\right)^2\right)\frac{1}{\pi_i\pi_j}
$$
where $I_{k\ne i}$ is an indicator variable equal to one when $k \ne i$ and zero otherwise, and then estimate $\tau_2$ with
$$
\hat\tau_2 = \frac{4}{N^2(N-1)^2}\left(\hat\tau_{2a} - \hat b_{\hat\tau_{2a}}\right).
$$

\begin{remark}
\label{rem:tau1negative}
As a consequence of the variability when estimating $\hat\tau_1$ and $\hat b_{\hat\tau_1}$, in rare instances, $\hat\tau_{1,BC}$ may be negative. A similar phenomenon may occur with $\hat\tau_2$ and $\hat b_{\hat\tau_{2a}}$. We thus impose a floor at zero and compute $\hat\tau_{1,BCF}=\max(\hat\tau_{1,BC},0)$ and $\hat\tau_{2,BCF}=\max(\hat\tau_2,0)$.
\end{remark}

Putting the two pieces together leads to our final estimator of $\Var(\hat a)$:
\begin{equation}
\label{eq:varhat}
\widehat\Var(\hat a) = 4\hat\tau_{1,BCF} + \hat\tau_{2,BCF}.
\end{equation}

\section{Case study: American Community Survey}
\label{sec:sims}

In this section, we demonstrate the use of our estimator defined in Equation \eqref{eq:varhat} to estimate $\Var(\hat a)$, and contrast its performance against that of the classical asymptotic estimator given in Equation \eqref{eq:asymptotic}. Our goal is to explore the performance of these estimators across sample sizes, sampling designs, and outcome distributions. Notably, we are interested in studying performance beyond Poisson sampling schemes. 

Code to reproduce all results in this section is available at  \if0\blind
{\url{https://github.com/AmeerD/UVvariances}}\fi \if1\blind{XXX}\fi.

\subsection{Context and setup}

To replicate a practical use case, we will study the 2022 1-year American Community Survey (ACS). The ACS is an annual household survey conducted by the US Census Bureau that provides an ongoing assessment of the economic, demographic, and social state of the United States \citep{ACS}. We collect the $N=39,821$ respondents who are members of the labour force residing in Washington state with positive income; the data are accessed using the \texttt{tidycensus} R package \citep{tidycensus}. These respondents form our fixed finite population of interest. From this population, we draw $2,500$ samples under each of the following $23$ sampling designs:
\begin{itemize}
    \item \textbf{Bernoulli}: Each unit is sampled independently of other units and has an equal probability of inclusion. The expected number of sampled units is set to $E[n]=30,50,100,250,500$.
    \item \textbf{Poisson}: Each unit is sampled independently of other units and has an inclusion probability proportional to the inverse of the average of its ACS sampling weight and the mean ACS sampling weight. The expected number of sampled units is set to $E[n]=30,50,100,250,500$.
    \item \textbf{SRSWOR}: Units are sampled using simple random sampling without replacement with $n=30,50,100,\allowbreak250,500$.
    \item \textbf{Stratified}: Treating the $61$ public use microdata areas (PUMAs) in Washington as strata, we draw $n_s=2,5,9$ units from each strata using simple random sampling without replacement for a total sample size of $n=122,305,549$, respectively.
    \item \textbf{Two-stage cluster sampling}: Treating the $61$ PUMAs as clusters, we first sample a subset of clusters, and then within each sampled cluster, draw a subset of units with simple random sampling without replacement. The number of clusters sampled is $n_c=15,25,50,50,50$ and the number of units sampled per cluster is $n_u=2,2,2,5,10$ for a total sample size of $n=30,50,100,250,500$, respectively.
\end{itemize}

Our goal is to estimate the variance of the GREG estimator of the average total income for individuals, $y_i$, from these samples using age, travel time to work, and usual hours worked per week as covariates, $x_i$. We examine income on the original (i.e., dollars) and log-scale as we are interested in verifying the suitability of our variance estimator under multiple outcome distributions.

For each sample in each setting, we compute the GREG estimate of the average total person's income, the estimate of the asymptotic variance using $\eqref{eq:asymptotic}$, and the H-decomposition-based estimate of the exact variance using \eqref{eq:varhat}. As noted in Remark \ref{rem:ij}, \cite{efron2014estimation} recently showed that the infinitesimal jackknife can be used to compute the asymptotic variance of U- and V-statistics: it targets $(k+1)^2\tau_1$ though has been noted to suffer from potentially significant upwards bias. As an additional comparison point, we compute the infinitesimal jackknife estimate of the variance using the balanced variance estimation method (BM) \citep{zhou2021v}; see Supplement \ref{app:IJ} for details. These quantities are denoted by $\hat a^{(r)}$, $\hat v_{asy}^{(r)}$, $\hat v_{HD}^{(r)}$, and $\hat v_{IJ}^{(r)}$, respectively, where $r=1,\dots,2500$ indexes the samples. 

We assess performance in each setting using two metrics. First, let $\hat v_{emp}=\frac{1}{2500-1}\sum_{r=1}^{2500}\left(\hat a^{(r)}-\bar{\hat a}\right)^2$ denote the empirical variance of the GREG estimates where $\bar{\hat a}=\frac{1}{2500}\sum_{r=1}^{2500}\hat a^{(r)}$ is the empirical mean. The two metrics are then: 
\begin{enumerate}
    \item Variance ratios: For $r=1,\dots,2500$, the variances ratios for the asymptotic estimator, our H-decomposition-based estimator, and the infinitesimal jackknife estimator are given by $\hat v_{asy}^{(r)}/\hat v_{emp}$, $\hat v_{HD}^{(r)}/\hat v_{emp}$, and $\hat v_{IJ}^{(r)}/\hat v_{emp}$, respectively. Distributions of variance ratios centred and concentrating at $1$ are preferred.
    \item Empirical coverage: We compute coverage of the confidence intervals for $a=\frac{1}{N}\sum_{i=1}^{39,281}y_i$ at the $\alpha=0.8,0.9,0.95$ levels. For $r=1,\dots,2500$, we first compute $(100\alpha)\%$ confidence intervals with $\hat a^{(r)} \pm q_{(1-\alpha)/2}\sqrt{\hat v_{asy}^{(r)}}$, $\hat a^{(r)} \pm q_{(1-\alpha)/2}\sqrt{\hat v_{HD}^{(r)}}$, $\hat a^{(r)} \pm q_{(1-\alpha)/2}\sqrt{\hat v_{IJ}^{(r)}}$, or $\hat a^{(r)} \pm q_{(1-\alpha)/2}\sqrt{\hat v_{emp}}$ for the asymptotic, H-decomposition-based, infinitesimal jackknife, and empirical variance estimators, respectively, where $q_{(1-\alpha)/2}$ is the $(1-\alpha)/2$ quantile of the standard Gaussian distribution. Coverage is then computed for each variance estimator by counting the proportion of intervals that contain $a$.
\end{enumerate}

\subsection{Results}

\begin{figure}[htp]
\centering
\includegraphics[width=1\textwidth]{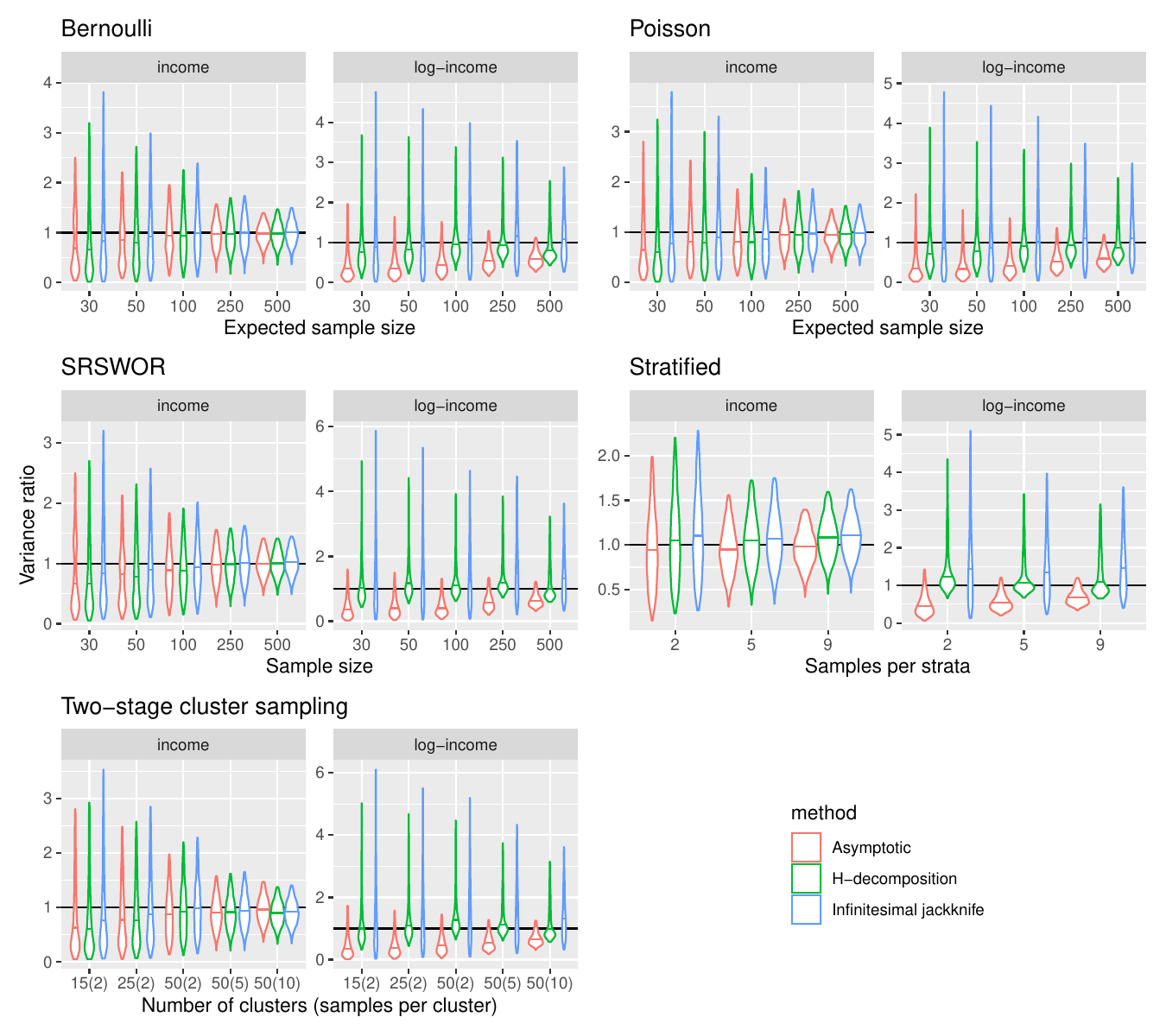}
\caption{Violin plots of the empirical distribution of the 2,500 variance ratios computed for all three variance estimation methods, truncated at the 95\% quantile. Each panel corresponds to a sampling design and an outcome variable. Variance ratios near one are preferred. All three estimators perform well when income is the response variable. For log-income, however, our H-decomposition-based estimator performs best: the asymptotic estimator suffers from severe downwards bias and the infinitesimal jackknife has higher variance.}
\label{fig:vrtrunc}
\end{figure}

Figure \ref{fig:vrtrunc} plots the distributions of the variance ratios for all three estimators in all settings considered. Each panel corresponds to a sampling design and an outcome variable (income or log-income). Within each panel, there is one set of violin plots for each sample size setting and median estimates are indicated with a horizontal line within the corresponding violin. The three variance estimation methods are indicated by colour. Notice that the distributions are all right-skewed. This is expected: as variances are positive and based on sums of squares, they can be unstable in small samples. Occasionally, all three methods produce large outlier estimates which can obscure the bulk of the distribution when plotting. Here we have truncated the distributions at the 95\% quantile to facilitate visualisation of the modes of the variance ratio distributions, and provide a second set of plots without truncation in Supplement \ref{app:sims}. 

Focusing first on the original-scale income plots, under all designs, all three methods perform quite well: the distributions increasingly concentrate near one as sample sizes increase, though in the smallest sample sizes, all methods tend to underestimate slightly. The log-income plots are more interesting. Under all five designs, the asymptotic estimator grossly underestimates the variance. While this issue moderates as the sample size increases, it is not fully resolved in the sample sizes we consider. By contrast, the two U- and V-statistic based methods do not suffer from such downwards bias. Indeed, our H-decomposition-based estimator is centred at or near one in all settings. The infinitesimal jackknife also outperforms the asymptotic estimator, but tends to be more variable than our approach, and in some cases, such as under the stratified design, seems to be biased upwards. This is a well known deficiency of the infinitesimal jackknife; see \cite{zhou2021v} for a discussion. 

Our variance estimator was designed under the assumption of Poisson sampling as the independence of the sampling indicators leads to a meaningful simplification. We rationalize this decision by suggesting that for non-independent designs, the discarded terms will be small. The results in Figure \ref{fig:vrtrunc} suggest that this is reasonable in practice as our estimator does not appear to experience any meaningful degradation in performance under any of the non-independent designs.

\begin{figure}[H]
\centering
\includegraphics[width=1\textwidth]{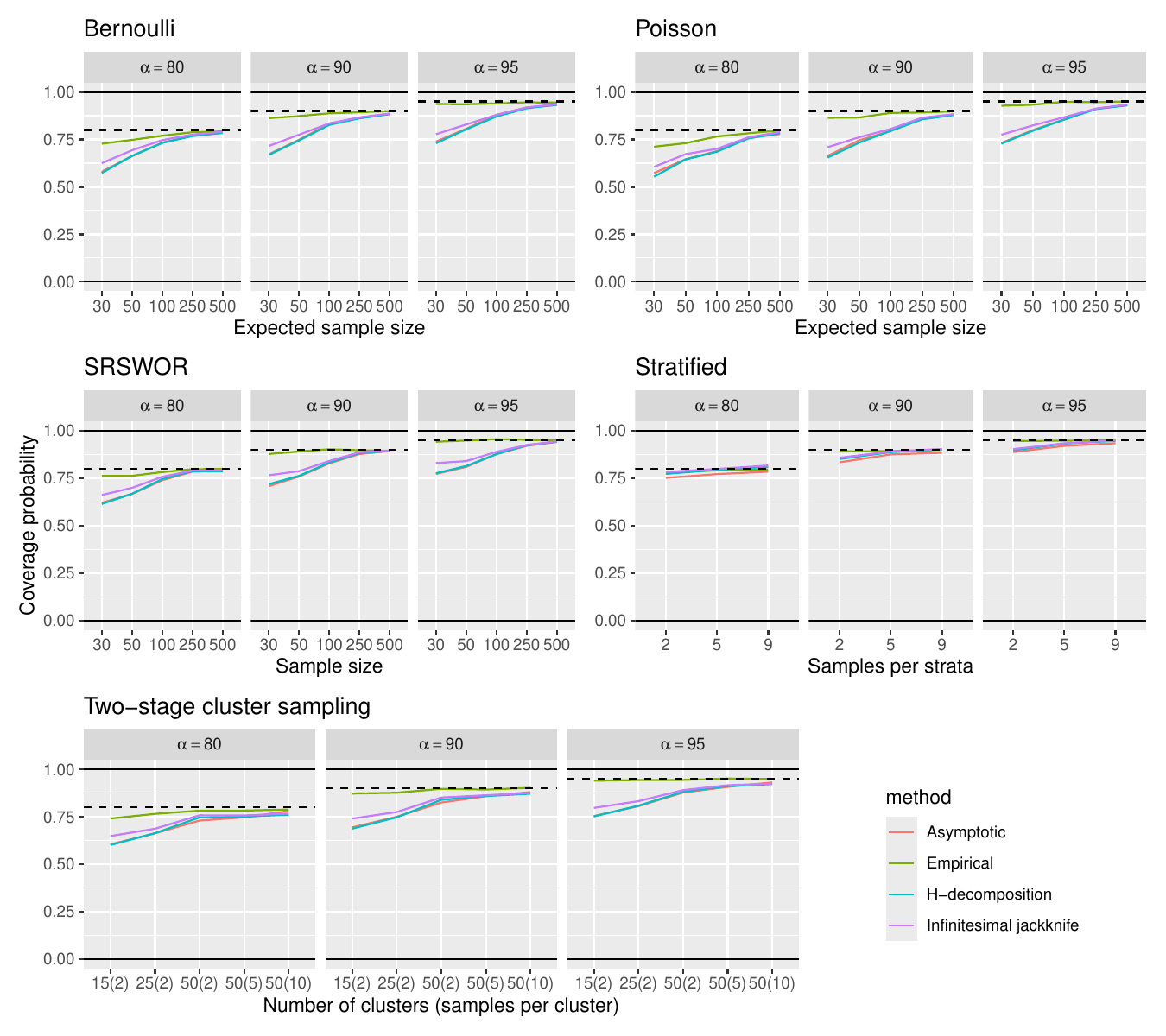}
\caption{Empirical coverage curves for the original-scale income analysis. Each panel corresponds to a sampling design and nominal coverage level. All three estimators perform similarly, undercovering in small sample sizes and approaching the nominal level as the sample size increases.}
\label{fig:covreal}
\end{figure}

\begin{figure}[H]
\centering
\includegraphics[width=1\textwidth]{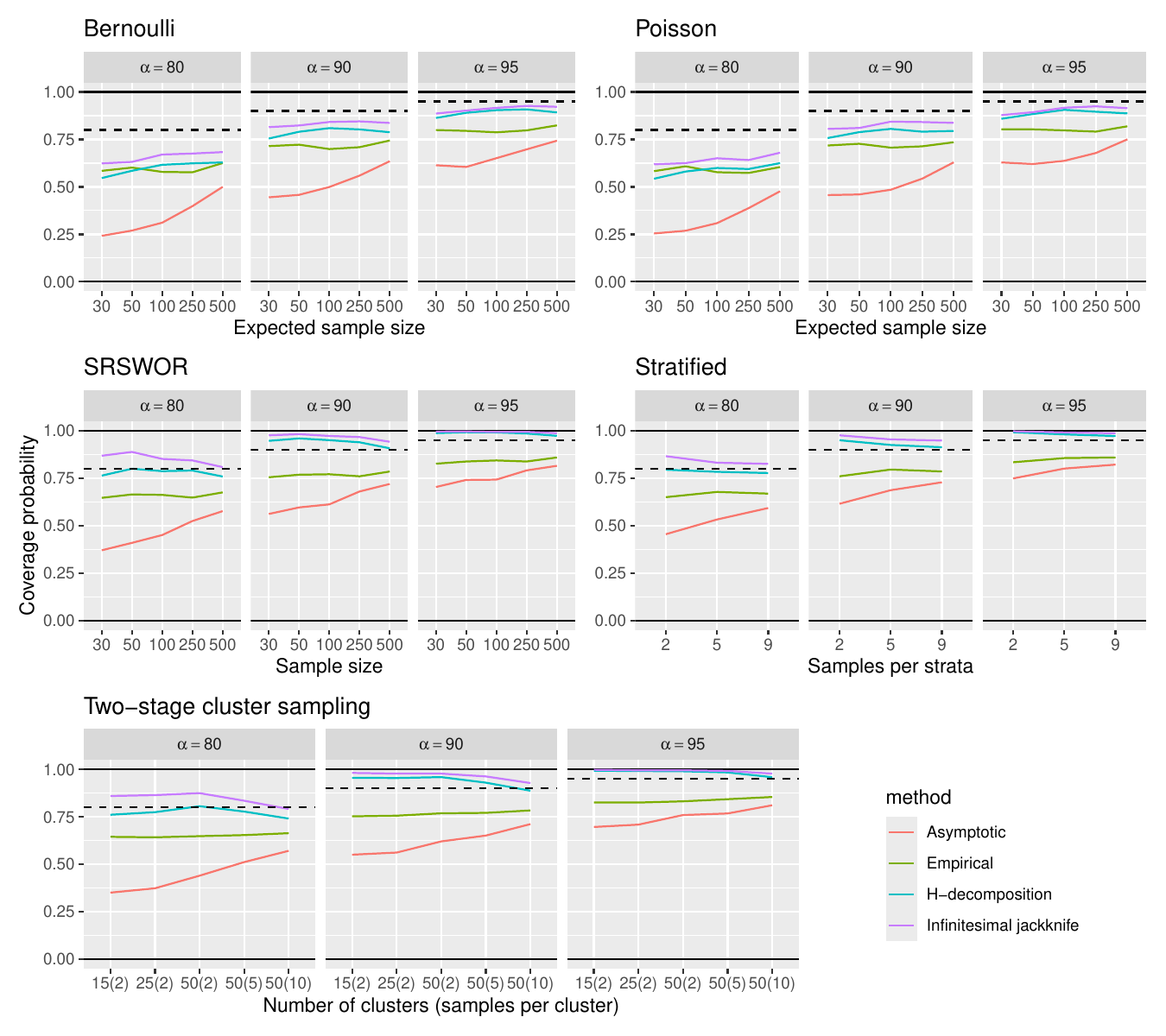}
\caption{Empirical coverage curves for the log-scale income analysis. Each panel corresponds to a sampling design and nominal coverage level. Confidence intervals derived from the asymptotic estimator suffer from severe undercoverage. Under Bernoulli and Poisson sampling, the H-decomposition approach performs similarly to the intervals constructed using the empirical variance. Under non-independent designs, both the H-decomposition and infinitesimal jackknife variance estimators produce conservative confidence intervals.}
\label{fig:covlog}
\end{figure}

Figures \ref{fig:covreal} and \ref{fig:covlog} display the empirical coverage results for the analysis using income and log-income as the outcome variables, respectively. Each figure contains three panels per sampling design, one for each of the three $\alpha$ levels, and each panel plots coverage as a function of sample size. Similar to the variance ratio results, in Figure \ref{fig:covreal}, we see little difference in performance across variance estimation methods, with coverage improving as sample size increases. This mimics the findings of Figure \ref{fig:vrtrunc}. By contrast, in Figure \ref{fig:covlog}, the intervals based on the asymptotic variance estimator severely undercover whereas the H-decomposition and infinitesimal jackknife estimators are closer to the empirical variance coverage curves and the nominal levels. We note that under non-independent designs, the H-decomposition and infinitesimal jackknife estimators tend to overcover relative to the empirical variance confidence intervals, suggesting that the omitted covariance terms lead to conservative confidence intervals. We comment that in the log-income analysis, confidence intervals constructed using the empirical variance estimator are not achieving the nominal coverage level, especially in smaller sample sizes, suggesting that asymptotic normality does not hold. We verify that this is indeed the case in Supplement \ref{app:sims}. Thus, care must be taken in making inferential statements in such contexts, even when more reliable variance estimates can be constructed. 

We conclude this section with a brief assessment of running times for each algorithm. We record the wall clock running time for each variance computation. In short, all three methods are fast: the vast majority of computations require under one second, and the highest running time for our approach is $4.3-12.1$ seconds, occurring  when the (expected) sample sizes are 500. Comparing the three estimators, the asymptotic estimator is the fastest, followed by the infinitesimal jackknife, followed by the H-decomposition-based approach. This is not surprising as these estimators are of increasing complexity. The full range of running times are summarised in Table \ref{tab:runtime} in Supplement \ref{app:sims}.

\section{Discussion}
\label{sec:discussion}

Reliable variance estimation is a constant challenge in modern survey statistics: as the complexity of estimators increases, so does the difficulty of variance estimation. In the context of model-assisted estimation, the assumptions underpinning the classical asymptotic variance estimator are often untenable in practice; if the uncertainty originating from the working model is non-negligible, it will underestimate the variance. In this paper, we propose a different approach. We suggest exploiting structural properties of the working model in order to derive estimators of the exact variance. Specifically, when the working model admits a U- or V-statistic representation, we demonstrate that this structure propagates to the model-assisted estimator. This fact unlocks exact variance expressions based on the corresponding H-decomposition. Through a comprehensive study of the generalized regression estimator, we then demonstrate the utility of our estimator and its superior performance relative to the asymptotic approach. 

Challenges in variance estimation are compounded as survey designs become increasingly complex. The exact variance expression derived in \eqref{eq:Hvar} is valid regardless of the sampling design, though estimating every term remains difficult for highly complex surveys. In Section \ref{sec:varest}, we build our estimator of $\Var(\hat a)$ for Poisson sampling designs as this leads to an elegant simplification of the exact variance. In Section \ref{sec:sims}, we then explore the empirical performance of the estimator in several non-independent designs and find that it performs well. In future work, it would be interesting to formally assess how far one can deviate from Poisson sampling before the performance of our estimator degrades beyond that of the asymptotic estimator.

The class of U- and V-statistics is wide. In the introduction, we comment that it includes simple methods such as the linear model as well as more complicated ensemble models. We briefly expand on the latter here. Recent work has demonstrated that ensemble models, most famously the random forests of \cite{breiman1996bagging} and \cite{breiman2001random}, are by definition either a U- or V-statistic, depending on whether samples are drawn without or with replacement, respectively \citep{wager2014confidence, mentch2016quantifying, wager2018estimation, zhou2021v}. Future work may focus on developing efficient exact variance estimation algorithms for such models that additionally account for practical issues such as the large number of terms in \eqref{eq:Hvar}, external randomness in the kernel, and Monte Carlo variation stemming from the incomplete nature of such models. Looking further, it will be of great interest to enumerate the entire set of predictive models that admit U- and V-statistic representations to understand the full extent to which our approach can be applied.

The difference estimator is not unique to survey statistics and enjoys widespread usage in several other areas, though often under different names. Some examples are the augmented inverse probability weighted (AIPW) estimator in causal inference \citep{robins1994estimation}, and the recent prediction-powered inference \citep{angelopoulos2023prediction} and active inference \citep{zrnic2024active} estimators from statistical learning. It thus immediately follows that our ideas may find use in those areas for exact variance estimation, though domain specific challenges remain, such as understanding how modelled weights affect the validity of the U- and V-statistic representations.

Our paper is about exact variance estimation, though we suspect there are consequences for the asymptotic study of model-assisted and related estimators. Some questions include whether one can express the classical asymptotic variance estimator using the terms in \eqref{eq:Hvar}, and if so, whether that can be used to devise a general second-order correction. Contemporary work in the dependent U- and V-statistics literature seeks to prove asymptotic properties by identifying conditions for which the asymptotic distribution of a U- or V-statistic constructed with dependent data approaches the analogous independent case \citep{dehling2006limit, dehling2010central, lee2019u}. We, by contrast, study model-assisted estimators that are known to be asymptotically normal (using the superpopulation framework; see \citealt{breidt2017review} for a review) under designs that induce data dependence, and find hidden U- and V-statistic structure. It is of interest to understand the connections between these approaches, and explore whether our work can be used to identify new conditions with respect to dependence structure, properties of the kernels, or otherwise.

\section*{Acknowledgments}

Ameer Dharamshi was supported by the Natural Sciences and Engineering Research Council of Canada. Jon Wakefield was supported by R37 AI029168 and R01 HD112421 from the National Institutes of Health of the United States. 

\bibliographystyle{natbib}
\bibliography{UVGREG}

\newpage

\appendix
\section{Proofs}
\label{app:proofs}

\subsection{Proof of Lemma \ref{lem:popvstat}}

\begin{proof}
Observe that:
\begin{align*}
\hat m(x_0)&=\frac{1}{n^k}\sum_{j_1\in S}\dots\sum_{j_k\in S}h(x_0|z_{j_1},\dots,z_{j_k})\\
&=\frac{1}{N^k}\sum_{j_1\in S}\dots\sum_{j_k\in S}\frac{N^k}{n^k}h(x_0|z_{j_1},\dots,z_{j_k})\\
&=\frac{1}{N^k}\sum_{j_1\in U}\dots\sum_{j_k\in U}\frac{N^k}{n^k}h(x_0|z_{j_1},\dots,z_{j_k})\prod_{l=1}^k I_{j_l}
\end{align*}
\end{proof}

\subsection{Proof of Theorem \ref{thm:MAEvstat}}

\begin{proof}
Observe that $\hat a$ can be re-written as:
\begin{align*}
\hat a &= \frac{1}{N}\sum_{i\in U} \hat m(x_i) + \frac{1}{N}\sum_{i\in S}\frac{y_i- \hat m(x_i)}{\pi_i} \\
&= \frac{1}{N}\sum_{i\in U} \left[\frac{y_iI_i}{\pi_i} + \left(1-\frac{I_i}{\pi_i}\right)\hat m(x_i)\right] \\
&= \frac{1}{N}\sum_{i\in U} \left[\frac{y_iI_i}{\pi_i} + \left(1-\frac{I_i}{\pi_i}\right)\frac{1}{N^k}\sum_{j_1\in U}\dots\sum_{j_k\in U}\frac{N^k}{n^k}h(x_i|z_{j_1},\dots,z_{j_k})\prod_{l=1}^k I_{j_l}\right] \\
&= \frac{1}{N}\sum_{i\in U} \frac{1}{N^k}\sum_{j_1\in U}\dots\sum_{j_k\in U}\left[\frac{y_iI_i}{\pi_i} + \left(1-\frac{I_i}{\pi_i}\right)\frac{N^k}{n^k}h(x_i|z_{j_1},\dots,z_{j_k})\prod_{l=1}^k I_{j_l}\right] \\
&= \frac{1}{N^{k+1}}\sum_{i\in U} \sum_{j_1\in U}\dots\sum_{j_k\in U}\left[\frac{y_iI_i}{\pi_i} + \left(1-\frac{I_i}{\pi_i}\right)\frac{N^k}{n^k}h(x_i|z_{j_1},\dots,z_{j_k})\prod_{l=1}^k I_{j_l}\right].
\end{align*}
The above resembles a V-statistic, though the expression inside the summations is not symmetric in its arguments. Replacing the indexing label $i$ with $j_{k+1}$, we can symmetrize the kernel to yield our final result:
\begin{align*}
\hat a &= \frac{1}{N^{k+1}}\sum_{j_1\in U}\dots\sum_{j_{k+1}\in U}\left[\frac{y_{j_{k+1}}I_{j_{k+1}}}{\pi_{j_{k+1}}} + \left(1-\frac{I_{j_{k+1}}}{\pi_{j_{k+1}}}\right)\frac{N^k}{n^k}h(x_{j_{k+1}}|z_{j_1},\dots,z_{j_k})\prod_{l=1}^k I_{j_l}\right] \\
&= \frac{1}{N^{k+1}}\sum_{j_1\in U}\dots\sum_{j_{k+1}\in U}\frac{1}{k+1}\sum_{p=1}^{k+1}\left[\frac{y_{j_p}I_{j_p}}{\pi_{j_p}} + \left(1-\frac{I_{j_p}}{\pi_{j_p}}\right)\frac{N^k}{n^k}h(x_{j_p}|\{z_{j_1},\dots,z_{j_{k+1}}\}\setminus \{z_{j_p}\})\prod_{l\in\{1,\dots,k+1\}\setminus \{p\}} I_{j_l}\right].
\end{align*}
\end{proof}

\subsection{Proof of Proposition \ref{prop:Vgreg}}

\begin{proof}
We have that
\begin{align*}
\hat a &= \frac{1}{N}\sum_{i\in U} \left[\frac{y_iI_i}{\pi_i} + \left(1-\frac{I_i}{\pi_i}\right)\hat m(x_i)\right] \\
&= \frac{1}{N}\sum_{i\in U} \left[\frac{y_iI_i}{\pi_i} + \left(1-\frac{I_i}{\pi_i}\right)x_i^\top Q\sum_{j\in S} \frac{x_j y_j}{\pi_j}\right] \\
&= \frac{1}{N}\sum_{i\in U} \frac{y_iI_i}{\pi_i} + \frac{1}{N}\sum_{i\in U}\left[\left(1-\frac{I_i}{\pi_i}\right)x_i^\top Q\sum_{j\in S} \frac{x_j y_j}{\pi_j}\right] \\
&= \frac{1}{N}\sum_{i\in U} \frac{y_iI_i}{\pi_i} + \frac{1}{N}\sum_{i\in U}\left[\left(1-\frac{I_i}{\pi_i}\right)x_i^\top\right] Q\sum_{j\in U} \frac{x_j y_j I_j}{\pi_j} \\
&= \frac{1}{N}\sum_{i\in U} \frac{y_iI_i}{\pi_i} + \frac{1}{N}\left(t_x^\top - \sum_{i\in U}\frac{I_i}{\pi_i}x_i^\top\right) Q\sum_{j\in U} \frac{x_j y_j I_j}{\pi_j} \\
&= \frac{1}{N}\sum_{i\in U} \frac{y_iI_i}{\pi_i} + \frac{1}{N}\left(t_x^\top - \sum_{j\in U}\frac{I_j}{\pi_j}x_j^\top\right) Q\sum_{i\in U} \frac{x_i y_i I_i}{\pi_i} \\
&= \frac{1}{N}\sum_{i\in U}\left[ \frac{y_iI_i}{\pi_i} + \left(t_x^\top - \sum_{j\in U}\frac{I_j}{\pi_j}x_j^\top\right) Q\frac{x_i y_i I_i}{\pi_i}\right] \\
&= \frac{1}{N}\sum_{i\in U}\left[1 + \left(t_x^\top - \sum_{j\in U}\frac{I_j}{\pi_j}x_j^\top\right) Qx_i\right]\frac{y_iI_i}{\pi_i} \\
&= \frac{1}{N}\sum_{i\in U}\left[1+\sum_{j\in U}\left(\frac{1}{N}t_x^\top-\frac{I_j}{\pi_j}x_j^\top\right)Qx_i\right]\frac{y_iI_i}{\pi_i} \\
&= \frac{1}{N}\sum_{i\in U}\sum_{j\in U}\left[\frac{1}{N}+\left(\frac{1}{N}t_x^\top-\frac{I_j}{\pi_j}x_j^\top\right)Qx_i\right]\frac{y_iI_i}{\pi_i} \\
&= \frac{1}{N^2}\sum_{i\in U}\sum_{j\in U}\left[1+N\left(\frac{1}{N}t_x^\top-\frac{I_j}{\pi_j}x_j^\top\right)Qx_i\right]\frac{y_iI_i}{\pi_i} \\
&= \frac{1}{N^2}\sum_{i\in U}\sum_{j\in U}\frac{1}{2}\left[\left(1+N\left(\frac{1}{N}t_x^\top-\frac{I_j}{\pi_j}x_j^\top\right)Qx_i\right)\frac{y_iI_i}{\pi_i} + \left(1+N\left(\frac{1}{N}t_x^\top-\frac{I_i}{\pi_i}x_i^\top\right)Qx_j\right)\frac{y_jI_j}{\pi_j}\right],
\end{align*}
which is the form of a degree-$2$ V-statistic with kernel
$$
h^*(z_i,z_j)=\frac{1}{2}\left[\left(1+N\left(\frac{1}{N}t_x^\top-\frac{I_j}{\pi_j}x_j^\top\right)Qx_i\right)\frac{y_iI_i}{\pi_i}+\left(1+N\left(\frac{1}{N}t_x^\top-\frac{I_i}{\pi_i}x_i^\top\right)Qx_j\right)\frac{y_jI_j}{\pi_j}\right]
$$
as required.
\end{proof}

\subsection{Proof of Proposition \ref{prop:tau1bias}}

\begin{proof}
First, consider the following expansion of $\hat\tau_1$:
\begin{align*}
\hat\tau_1 &= \frac{1}{N^2}\sum_{i\in S}\left(\hat\phi_{1;i}'(I_i=1)-\hat\theta'_{1;i}\right)^2\frac{1}{1-\pi_i} \\ 
&= \frac{1}{N^2}\sum_{i\in S}\left(\phi_{1;i}'(I_i=1)-\theta'_{1;i}+\hat\phi_{1;i}'(I_i=1)-\hat\theta'_{1;i}-\phi_{1;i}'(I_i=1)+\theta'_{1;i}\right)^2\frac{1}{1-\pi_i} \\ 
&= \frac{1}{N^2}\sum_{i\in S}\left[\left(\phi_{1;i}'(I_i=1)-\theta'_{1;i}\right)^2+\left(\hat\phi_{1;i}'(I_i=1)-\hat\theta'_{1;i}-\phi_{1;i}'(I_i=1)+\theta'_{1;i}\right)^2\right.\\
&\left.\quad\quad\quad\quad\quad+2\left(\phi_{1;i}'(I_i=1)-\theta'_{1;i}\right)\left(\hat\phi_{1;i}'(I_i=1)-\hat\theta'_{1;i}-\phi_{1;i}'(I_i=1)+\theta'_{1;i}\right)\right]\frac{1}{1-\pi_i} \\
&= \frac{1}{N^2}\sum_{i=1}^N\left[\left(\phi_{1;i}'(I_i=1)-\theta'_{1;i}\right)^2+\left(\hat\phi_{1;i}'(I_i=1)-\hat\theta'_{1;i}-\phi_{1;i}'(I_i=1)+\theta'_{1;i}\right)^2\right.\\
&\left.\quad\quad\quad\quad\quad+2\left(\phi_{1;i}'(I_i=1)-\theta'_{1;i}\right)\left(\hat\phi_{1;i}'(I_i=1)-\hat\theta'_{1;i}-\phi_{1;i}'(I_i=1)+\theta'_{1;i}\right)\right]\frac{I_i}{\pi_i}\cdot\frac{\pi_i}{1-\pi_i}.
\end{align*}

Next, take expectations with respect to the sampling indicators:
\begin{align*}
\E\left[\hat\tau_1\right]
&= \E\left[\frac{1}{N^2}\sum_{i=1}^N\left(\phi_{1;i}'(I_i=1)-\theta'_{1;i}\right)^2\frac{I_i}{\pi_i}\cdot\frac{\pi_i}{1-\pi_i}\right] \\
&\quad + \E\left[\frac{1}{N^2}\sum_{i=1}^N\left(\hat\phi_{1;i}'(I_i=1)-\hat\theta'_{1;i}-\phi_{1;i}'(I_i=1)+\theta'_{1;i}\right)^2\frac{I_i}{\pi_i}\cdot\frac{\pi_i}{1-\pi_i}\right] \\
&\quad + \E\left[\frac{2}{N^2}\sum_{i=1}^N\left(\phi_{1;i}'(I_i=1)-\theta'_{1;i}\right)\left(\hat\phi_{1;i}'(I_i=1)-\hat\theta'_{1;i}-\phi_{1;i}'(I_i=1)+\theta'_{1;i}\right)\frac{I_i}{\pi_i}\cdot\frac{\pi_i}{1-\pi_i}\right] \\
&= \frac{1}{N^2}\sum_{i=1}^N\left(\phi_{1;i}'(I_i=1)-\theta'_{1;i}\right)^2\E\left[\frac{I_i}{\pi_i}\right]\frac{\pi_i}{1-\pi_i}\\
&\quad + \frac{1}{N^2}\sum_{i=1}^N\E\left[\left(\hat\phi_{1;i}'(I_i=1)-\hat\theta'_{1;i}-\phi_{1;i}'(I_i=1)+\theta'_{1;i}\right)^2\frac{I_i}{\pi_i}\right]\frac{\pi_i}{1-\pi_i} \\
&\quad + \frac{2}{N^2}\sum_{i=1}^N\left(\phi_{1;i}'(I_i=1)-\theta'_{1;i}\right)\E\left[\left(\hat\phi_{1;i}'(I_i=1)-\hat\theta'_{1;i}-\phi_{1;i}'(I_i=1)+\theta'_{1;i}\right)\frac{I_i}{\pi_i}\right]\frac{\pi_i}{1-\pi_i} \\
&= \frac{1}{N^2}\sum_{i=1}^N\left(\phi_{1;i}'(I_i=1)-\theta'_{1;i}\right)^2\frac{\pi_i}{1-\pi_i}\\
&\quad + \frac{1}{N^2}\sum_{i=1}^N\E\left[\left(\hat\phi_{1;i}'(I_i=1)-\hat\theta'_{1;i}-\phi_{1;i}'(I_i=1)+\theta'_{1;i}\right)^2\right]\E\left[\frac{I_i}{\pi_i}\right]\frac{\pi_i}{1-\pi_i} \\
&\quad + \frac{2}{N^2}\sum_{i=1}^N\left(\phi_{1;i}'(I_i=1)-\theta'_{1;i}\right)\E\left[\hat\phi_{1;i}'(I_i=1)-\hat\theta'_{1;i}-\phi_{1;i}'(I_i=1)+\theta'_{1;i}\right]\E\left[\frac{I_i}{\pi_i}\right]\frac{\pi_i}{1-\pi_i} \\
&= \tau_1 + \frac{1}{N^2}\sum_{i=1}^N\Var\left(\hat\phi_{1;i}'(I_i=1)-\hat\theta'_{1;i}\right)\frac{\pi_i}{1-\pi_i}
\end{align*}
where in the second last step we used the independence in the inclusion indicators (as we are operating under Poisson sampling) and in the final step we used the fact that $\E\left[\hat\phi_{1;i}(I_i=1)-\hat\theta'_{1;i}\right]=\phi_{1;i}'(I_i=1)-\theta'_{1;i}$. 

The second expression in the proposition follows from the variance formula for Horvitz-Thompson estimators under Poisson sampling:
\begin{align*}
b_{\hat\tau_1} &= \frac{1}{N^2}\sum_{i=1}^N\Var\left(\hat\phi_{1;i}'(I_i=1)-\hat\theta'_{1;i}\right)\frac{\pi_i}{1-\pi_i} \\
&= \frac{1}{N^2}\sum_{i=1}^N\Var\left(\frac{1}{N-\frac{1}{\pi_i}}\sum_{j\ne i}\left(\phi_{1;i,j}(I_i=1)-\theta_{i,j}\right)\frac{I_j}{\pi_j}\right)\frac{\pi_i}{1-\pi_i} \\
&= \frac{1}{N^2}\sum_{i=1}^N\left(\frac{1}{\left(N-\frac{1}{\pi_i}\right)^2}\sum_{j\ne i}\frac{\left(\phi_{1;i,j}(I_i=1)-\theta_{i,j}\right)^2}{\pi_j^2}\Var\left(I_j\right)\right)\frac{\pi_i}{1-\pi_i} \\
&=\frac{1}{N^2}\sum_{i=1}^N\left(\frac{1}{\left(N-\frac{1}{\pi_i}\right)^2}\sum_{j\ne i}\frac{1-\pi_j}{\pi_j}\cdot\left(\phi_{1;i,j}(I_i=1)-\theta_{i,j}\right)^2\right)\frac{\pi_i}{1-\pi_i}.
\end{align*}
\end{proof}

\subsection{Proof of Proposition \ref{prop:tau2}}

\begin{proof}
From the definition of $\tau_2$,
\begin{align*}
\tau_2 &= \Var(H_2) = \Var\left(\frac{1}{{N\choose2}}\sum_{1\le i<j\le N}h_{2;i,j}(z_i,z_j)\right) \\
&=\frac{4}{N^2(N-1)^2}\left[\sum_{1\le i<j\le N}\Var\left(h_{2;i,j}(z_i,z_j)\right) + \sum_{\substack{1\le i_1<j_2\le N, 1\le i_2<j_2\le N \\ \{i_1,j_1\}\ne\{i_2,j_2\}}}\Cov\left(h_{2;i_1,j_1}(z_{i_1},z_{j_1}),h_{2;i_2,j_2}(z_{i_2},z_{j_2})\right) \right] \\
&=\frac{4}{N^2(N-1)^2}\left[\sum_{1\le i<j\le N}\E\left[h_{2;i,j}(z_i,z_j)^2\right] + \sum_{\substack{1\le i_1<j_2\le N, 1\le i_2<j_2\le N \\ \{i_1,j_1\}\ne\{i_2,j_2\}}}\E\left[h_{2;i_1,j_1}(z_{i_1},z_{j_1})h_{2;i_2,j_2}(z_{i_2},z_{j_2})\right] \right]
\end{align*}
where in the last step we use the fact that $E[h_{2;i,j}(z_i,z_j)]=0$ for all $1\le i<j\le N$. 

Under the simplifying  assumption that the second summation is small, the first term reduces as
\begin{align*}
\tau_2 &\approx \frac{4}{N^2(N-1)^2}\sum_{1\le i<j\le N}\E\left[h_{2;i,j}(z_i,z_j)^2\right] \\
&=\frac{4}{N^2(N-1)^2}\sum_{1\le i<j\le N}\E\left[\left(h_U(z_i,z_j)-\left(\phi'_{1;i}(z_i)-\theta'_{1;i}\right)-\left(\phi'_{1;j}(z_j)-\theta'_{1;j}\right)-\theta_{i,j}\right)^2\right] \\
&=\frac{4}{N^2(N-1)^2}\sum_{1\le i<j\le N}\E\left[\left(\left(h_U(z_i,z_j)-\phi'_{1;i}(z_i)-\phi'_{1;j}(z_j)+a\right)-\left(\theta_{i,j}-\theta'_{1;i}-\theta'_{1;j}+a\right)\right)^2\right] \\
&=\frac{4}{N^2(N-1)^2}\sum_{1\le i<j\le N}\left(\E\left[\left(h_U(z_i,z_j)-\phi'_{1;i}(z_i)-\phi'_{1;j}(z_j)+a\right)^2\right]-\left(\theta_{i,j}-\theta'_{1;i}-\theta'_{1;j}+a\right)^2\right) \\
&=\frac{4}{N^2(N-1)^2}\left(\underbrace{\sum_{1\le i<j\le N}\E\left[\left(h_U(z_i,z_j)-\phi'_{1;i}(z_i)-\phi'_{1;j}(z_j)+a\right)^2\right]}_{\tau_{2a}}-\underbrace{\sum_{1\le i<j\le N}\left(\theta_{i,j}-\theta'_{1;i}-\theta'_{1;j}+a\right)^2}_{\tau_{2b}}\right).
\end{align*}

\end{proof}
\section{Estimating the second component of $\tau_2$}
\label{app:tau2b}


The $\tau_{2b}$ term cannot be estimated in the same way as $\tau_{2a}$ since it involves $\theta_{i,j}$, which can only be estimated when $i,j\in S$. We therefore suggest a weighted approach similar to the one used for $\hat\tau_1$ and estimate $\tau_{2b}$ with
$$
\hat\tau_{2b}=\sum_{\substack{1\le i < j \le N \\ i,j\in S}}\frac{\left(\theta_{i,j}-\hat\theta'_{1;i}-\hat\theta'_{1;j}+\hat a\right)^2}{\pi_i\pi_j}.
$$

As mentioned in Section \ref{sec:varest}, $\tau_{2b}$ is negligible relative to $\tau_1$ and $\tau_{2a}$. To verify this claim, we repeat the simulation study in Section \ref{sec:sims} with the following additional estimator that includes $\hat\tau_{2b}$:
\begin{equation}
\label{eq:varhatprime}
\widehat\Var(\hat a)' = 4\hat\tau_{1,BCF} + \hat\tau_{2,BCF}'.
\end{equation}
where
$$
\hat\tau_{2,BCF}' = \max\left(\frac{4}{N^2(N-1)^2}\left(\hat\tau_{2a} - \hat\tau_{2b}-\hat b_{\hat\tau_{2a}}\right),0\right).
$$

\begin{figure}
\centering
\includegraphics[width=1\textwidth]{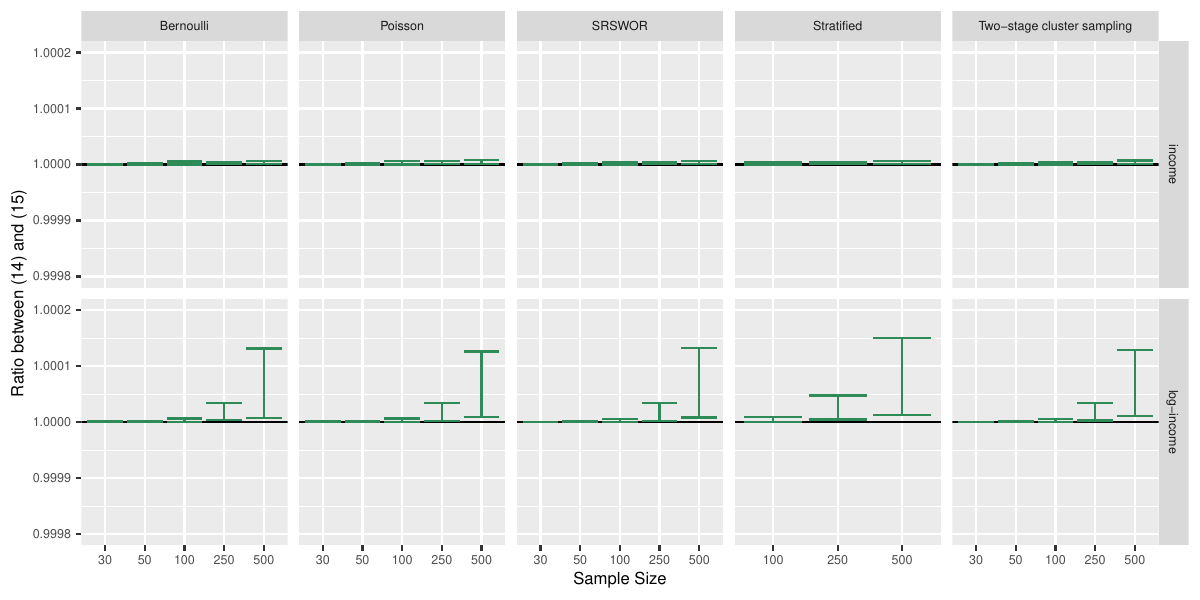}
\caption{Comparison of the H-decomposition-based variance estimators without and with $\tau_{2b}$. Each panel corresponds to a sampling design and an outcome variable. Within each panel, the range of ratios between the two estimators for each sample size are plotted using green lines. Note the scale of the plot. The exclusion of $\tau_{2b}$ leads to at most a 0.015\% increase in the variance estimate. Note: ``Sample Size" is used to refer to all of sample size (SRSWOR and two-stage cluster sampling), expected sample size (Bernoulli and Poisson), and total sample size (stratified sampling), rounded to conform to the other designs.}
\label{fig:tau2b}
\end{figure}

Figure \ref{fig:tau2b} plots the ratio between the variance estimator without $\tau_{2b}$ and the estimator with $\tau_{2b}$ (i.e. Equations \ref{eq:varhat} and \ref{eq:varhatprime}, respectively). No meaningful differences can be identified.

\section{Additional simulation results}
\label{app:sims}

\subsection{Additional variance ratio results}

\begin{figure}
\centering
\includegraphics[width=1\textwidth]{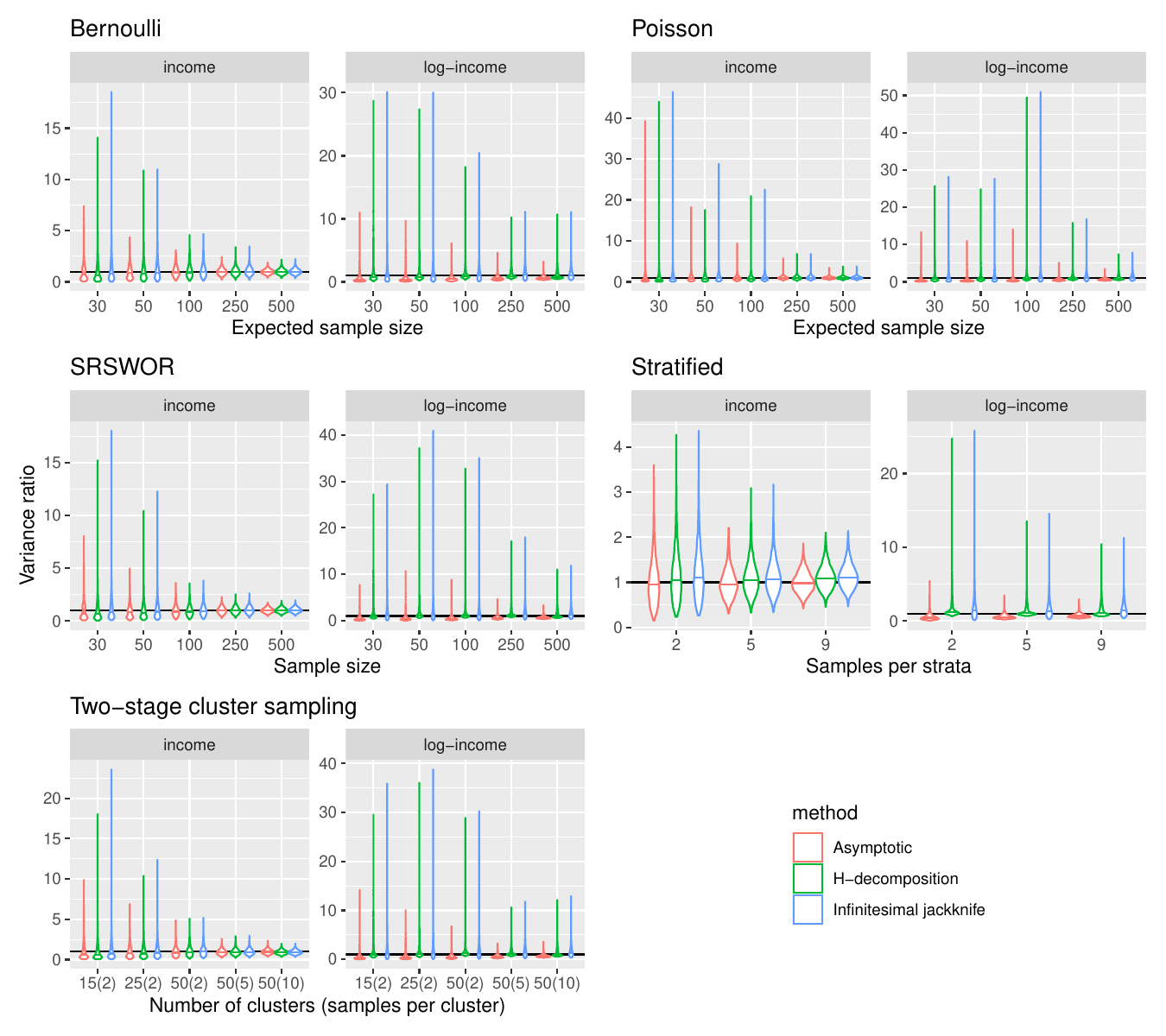}
\caption{Violin plots of the empirical distribution of the 2,500 variance ratios computed for all three variance estimation methods. Each panel corresponds to a sampling design and an outcome variable. Variance ratios near one are preferred.}
\label{fig:vrsupp}
\end{figure}

Figure \ref{fig:vrsupp} displays the full variance ratio distributions, without any truncation. Qualitatively, the plots are similar to Figure \ref{fig:vrtrunc} though condensed due to the presence of extreme outliers under all methods and designs when the sample size is small.

\subsection{Assessing normality}

\begin{figure}
\centering
\includegraphics[width=1\textwidth]{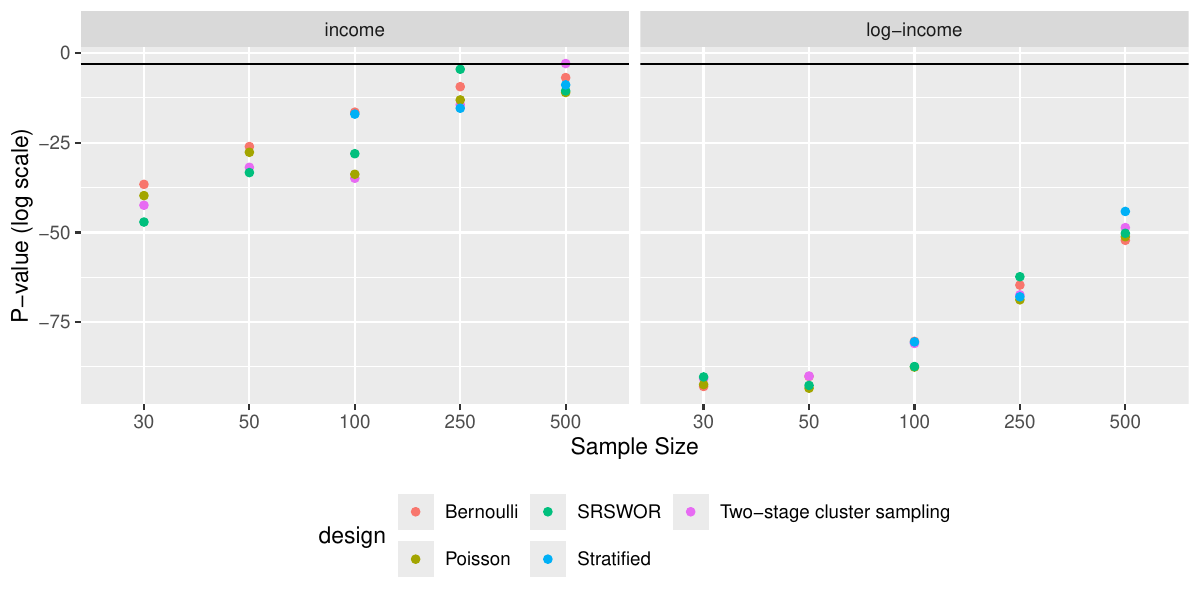}
\caption{Shapiro-Wilk test p-values computed for the 2,500 estimated $\hat a$ values under each sampling design and outcome setting. The p-values are all small (note that the values are plotted on the log scale), suggesting that the data are non-Gaussian. Note: ``Sample Size" is used to refer to all of sample size (SRSWOR and two-stage cluster sampling), expected sample size (Bernoulli and Poisson), and total sample size (stratified sampling), rounded to conform to the other designs.}
\label{fig:swtest}
\end{figure}

For each setting of the outcome distribution, sampling design, and sample size in Section \ref{sec:sims}, we conduct a Shapiro-Wilk test of the 2,500 estimates of $\hat a$ to assess whether the distributions are approximately Gaussian. Figure \ref{fig:swtest} plots the log of the p-values from these hypothesis tests along with a horizontal line at $\log(0.05)$.  The small p-values (excluding two instances of income on the original scale with sample sizes of 250 and 500) suggest that the distributions are non-Gaussian, clarifying why Gaussian confidence intervals generally fail to achieve the nominal levels in Figures \ref{fig:covreal} and \ref{fig:covlog}.

\subsection{Runtime analysis}
\label{app:runtime}

Table \ref{tab:runtime} contains wall-clock running time results for the simulation study conducted in Section \ref{sec:sims}. Since computation is invariant to sampling design and outcome distribution, we aggregate runtime results for all settings across these two dimensions. The column label ``Sample Size" is thus used to refer to all of sample size (SRSWOR and two-stage cluster sampling), expected sample size (Bernoulli and Poisson), and total sample size (stratified sampling), rounded to conform to the other designs. The table demonstrates that running times for the asymptotic method are the lowest and for the H-decomposition method are the highest. This is not unexpected: the H-decomposition variance estimator is far more complicated, require multiple iterations through the sample to compute its constituent components. However, the distinction is practically irrelevant. Even in the largest sample size of 500, the running time reaches a maximum of 12 seconds.

\begin{table}
\centering
\caption{Summary of variance estimation running times. Each cell contains the range of running times in seconds for the corresponding sample size* and estimation method, rounded to three decimal places. \\
*See text in Section \ref{app:runtime}.}
\label{tab:runtime}
\begin{tabular}{||r||c c c||}
\hline
Sample Size & Asymptotic & H-decomposition & Infinitesimal Jackknife\\
\hline
\hline
30 & 0-0.003 & 0.002-0.251 & 0-0.111\\
\hline
50 & 0-0.001 & 0.007-0.058 & 0-0.002\\
\hline
100 & 0-0.002 & 0.037-0.464 & 0.001-0.01\\
\hline
250 & 0-0.002 & 0.619-2.525 & 0.003-0.017\\
\hline
500 & 0.001-0.007 & 4.344-12.067 & 0.014-0.068\\
\hline
\end{tabular}
\end{table}
\section{Connections to the infinitesimal jackknife}
\label{app:IJ}

The infinitesimal jackknife is the limiting case of the jackknife variance estimator where for each replicate, instead of eliminating a given unit, we instead downweight this unit by an infinitesimal amount \citep{jaeckel1972IJ, efron1982jackknife}. \cite{efron2014estimation} ignited new interest in the infinitesimal jackknife by deriving a convenient expression for the infinitesimal jackknife estimator of the variance of bootstrap aggregated estimators, which are a class of (incomplete) V-statistics. Since then, several authors have expanded on the use of the infinitesimal jackknife to estimate the variance of U- and V-statistics, most notably to estimate the variance of predictions made by random forests (and related tree-based ensemble algorithms) in a variety of contexts \citep{wager2014confidence, mentch2016quantifying, wager2018estimation, zhou2021v}.

These previous works have focused on estimating the asymptotic variance of U- and V-statistics. By contrast, our interest is in the exact variance of $\hat a$. The connection lies in an insight in \cite{wager2014confidence}: assuming independent data, the infinitesimal jackknife is targeting the variance of the Hájek projection, which is the first term of the H-decomposition. That is, it targets $(k+1)^2\tau_1$.

For the remainder of this section, we focus on the case where $k+1=2$ to align with our discussion of GREG. Let $\delta_{l\in\{i,j\}}$ be an indicator equal to one if $l$ is in the set $\{i,j\}$ and zero otherwise. Then, the expression for the infinitesimal jackknife estimator (with the U-statistics correction proposed by \citealt{wager2018estimation}) is given by
\begin{align*}
\label{eq:IJ}
\hat\tau_{1,IJ} &= \frac{N(N-1)}{4(N-2)^2}\cdot\sum_{l=1}^N\widehat\Cov\left(\delta_{l\in\{i,j\}},h_U(z_i,z_j)\right)^2 \\
&= \frac{N(N-1)}{4(N-2)^2}\sum_{l=1}^N\left[\frac{1}{{N\choose2}}\sum_{1\le i < j \le N}\left(\delta_{l\in\{i,j\}}-\frac{N-1}{{N\choose2}}\right)\left(h_U(z_i,z_j)-\hat a\right)\right]^2 \\
&= \frac{1}{N(N-1)(N-2)^2}\sum_{l=1}^N\left[\sum_{1\le i < j \le N}\left(\delta_{l\in\{i,j\}}-\frac{2}{N}\right)\left(h_U(z_i,z_j)-\hat a\right)\right]^2.
\end{align*}

Directly computing $\hat\tau_{1,IJ}$ using the above expression is computationally challenging. \cite{zhou2021v} propose an alternative, the balanced variance estimation method (BM).  Briefly, for each unit in the population, BM computes the average of all kernels that include the unit (i.e. the empirical estimates of $\phi'_{1;i}(z_i)$), and then computes the sample variance of these averages. This method is identical to $\hat\tau_{1,IJ}$ (up to a re-scaling of $(N-1)/N$) when the target U- or V-statistic satisfies a ``balanced subsample structure", which is automatic for complete U-statistics like $\hat a$; see \citealt{zhou2021v} for technical details. Formally, 
$$
\hat\tau_{1,BM} = \frac{(N-1)^2}{N^2(N-2)^2}\sum_{i=1}^N\left(\hat\phi'_{1;i}(z_i)-\hat a\right)^2
$$
where $\hat\phi'_{1;i}(z_i)$ and $\hat a$ are as defined in Section \ref{sec:uv}, leading to the following variance estimator:
\begin{equation}
\label{eq:bm}
\widehat{\text{Var}}(\hat a) = 4\hat\tau_{1,BM}
\end{equation}

The utility of this approach over the original infinitesimal jackknife is that there is regularity in $\hat\phi'_{1;j}(z_i)$ based on whether $i,j\in S$ that can be exploited to accelerate computations. 

In Section \ref{sec:sims}, we implement \eqref{eq:bm} to compare against our proposal, though find that as expected, in small samples, the infinitesimal jackknife and its equivalents are highly variable and suffer from upwards bias \citep{wager2014confidence, zhou2021v}. While several authors have proposed bias corrections, we do not consider them here as they too tend to perform poorly in small or moderate sample sizes, leading to frequent negative variance estimates.

\end{document}